\documentclass[twocolumn]{aastex7}
\usepackage{float}
\usepackage{amsmath}
\usepackage{soul}
\usepackage{hyperref}

\newcommand{\appropto}{\mathrel{\vcenter{
  \offinterlineskip\halign{\hfil$##$\cr
    \propto\cr\noalign{\kern2pt}\sim\cr\noalign{\kern-2pt}}}}}

\newcommand{\msun}{\mathrm{M_{\odot}}} 

\newcommand{\mwd}{M_{\mathrm{WD}}} 

\newcommand{\mstar}{M_{\star}}
\newcommand{\rstar}{R_{\star}}

\newcommand{\Teff}{T_{\mathrm{eff}}}

\newcommand{\rsun}{\mathrm{R_{\odot}}} 

\newcommand{\iniMstar}{M^{i}_{\star}}

\newcommand{\iniPc}{P^{i}_{\mathrm{c}}}
\newcommand{\inirhoc}{\rho^{i}_{\mathrm{c}}}

\newcommand{\iniPavg}{P^{i}_{\mathrm{avg}}}

\newcommand{\iniSc}{s^{i}_{\rm c}}

\newcommand{\mesa}{{\tt\string MESA}}
\newcommand{\athena}{{\tt\string Athena++}}

\newcommand{\kB}{k_{\mathrm{B}}}
\newcommand{\NA}{N_{\mathrm{A}}}

\newcommand{\gcc}{\mathrm{g} \, \mathrm{cm}^{-3}}
\newcommand{\kms}{\mathrm{km}\,\mathrm{s}^{-1}}

\newcommand{\Porb}{P_{\mathrm{orb}}}

\newcommand{\rhoc}{\ensuremath{\rho_{\rm c}}}

\newcommand{\Pej}{P_{\mathrm{ej}}}
\newcommand{\Mej}{M_{\mathrm{ej}}}
\newcommand{\Eej}{E_{\mathrm{ej}}}
\newcommand{\Eke}{E_{\mathrm{KE}}}

\begin{document}

\title{Mass Loss and Subsequent Thermal Evolution of Surviving Helium White Dwarfs Shocked by Thermonuclear Supernovae}

\author[0000-0001-9195-7390,sname='Wong',gname='Tin Long Sunny']{Tin Long Sunny Wong}
\affiliation{Department of Physics, University of California, Santa Barbara, CA 93106, USA}
\email[show]{tinlongsunny@ucsb.edu}

\author[0000-0001-8038-6836]{Lars Bildsten}
\affiliation{Department of Physics, University of California, Santa Barbara, CA 93106, USA}
\affiliation{Kavli Institute for Theoretical Physics, University of California, Santa Barbara, CA 93106, USA}
\email[]{bildsten@kitp.ucsb.edu}

%


\begin{abstract}

Following a type Ia supernova (SN Ia) in a double white dwarf (WD) binary, a surviving WD companion leaves at its orbital velocity $\approx 1$,000--3,000 km/s. 
The Gaia mission has discovered seven such hypervelocity WDs with inflated radii indicative of shock heating by SN ejecta. 
We study the interaction between SN ejecta and Roche lobe-filling 0.08--0.45\,$M_{\odot}$ helium WD companions using three-dimensional hydrodynamical simulations with {\tt Athena++}. 
Given the importance of the later thermal evolution, we include an accurate equation-of-state for the degenerate helium WD donor. 
We show that a lower-mass, larger-radius WD companion is more strongly impacted by SN ejecta and undergoes substantial mass loss. We find a tight relation between the fractional mass loss and the ratio between the ejecta ram pressure and donor volume-averaged pressure, which can be used for predicting mass loss in other systems. In the most extreme case, the companion becomes a very inflated $\approx0.02\,M_{\odot}$ object. We find helium mass loss $\approx 0.005 - 0.06\,M_{\odot}$ with velocities $\approx $~1,000$-$4,000\,km\,s$^{-1}$, which may lead to emission lines in the nebular phase. The surviving helium WD receives a kick velocity, but its final velocity is essentially determined by its orbital velocity $\lesssim$~1,600\,km\,s$^{-1}$. 
We model the post-explosion evolution of the shock-heated companions using {\tt MESA}, and find reasonable agreement with the hypervelocity stars D6-2, J0546+0836, J1332-3541 \& SDSS J1637+3631. 
A surviving $\gtrsim 0.3\,M_{\odot}$ helium WD can be ruled out in SN1972E \& SN2011fe, and any surviving helium WD is likely ruled out in SN remnants 0509-67.5 \& SN1006.

\end{abstract}


\section{Introduction}

Type Ia supernovae (SNe Ia) are important distance indicators used for inferring cosmological parameters \citep[][]{Riess1998,Perlmutter1999}. 
Despite broad consensus that SNe Ia originate from thermonuclear explosions of carbon-oxygen white dwarfs (CO WDs), the explosion mechanism as well as the progenitor systems remain debated \citep[see][for recent reviews]{Liu2023_review,Ruiter2025}. 

In the double detonation scenario, 
a He detonation wave is launched in the accreted He layer on a CO WD, sends a converging shock wave into the CO core, and subsequently triggers its detonation. 
The donor can be a He star in the single-degenerate scenario, or a He WD or another CO WD in the double-degenerate scenario. 
A tell-tale signature of the double detonation, double degenerate scenario is a surviving donor leaving at roughly its orbital velocity $\gtrsim 1$,000~$\kms$ once the binary unbinds. 
The discovery of 8 such hypervelocity stars from Gaia DR2 \& 3 and the Sloan Digital Sky Survey \citep[][]{Shen2018_D6,ElBadry2023,Hollands2025} strongly corroborates that this happens. 
Furthermore, these surviving WD donors show inflated radii and surface metal pollution, indicating interaction with SN ejecta. 
However, the interaction between the surviving donor and the SN ejecta, the surviving donor's subsequent stellar evolution, the relative contribution of this channel to SNe Ia, etc., remain relatively unexplored. It is not even clear whether the donor survives or detonates \citep[][]{Papish2015,Tanikawa2019,Boos2024}. 

Among these stars, D6-2 stands out as having the lowest velocity $\approx 1$,000~$\kms$ and the only one that traces back to a SN remnant \citep[age $\approx 10^{5}$~yrs;][]{Shen2018_D6}. Assuming a Roche lobe-filling WD donor, D6-2's low velocity suggests a wide orbit and hence low donor mass $\approx 0.1-0.2\,\msun$ \citep[][]{Bauer2021_D6}, which points to its origin as a low-mass He WD. 

Such He WDs naturally arise from AM CVn binaries, which are ultracompact binaries with orbital periods $\lesssim 1$~hr undergoing He mass transfer, with a semi-degenerate, $\lesssim 0.1\,\msun$ He-rich donor \citep[e.g.,][]{Ramsay2018,Green2025}. 
In the AM CVn Last Flash scenario \citep[][]{Bildsten2007}, the accretor undergoes a series of He flashes with increasing strength as the mass transfer rate $\dot{M}$ declines and the accumulated He mass increases. The accretor eventually undergoes one final, strongest flash, possibly leading to a double-detonation SN Ia. A double-detonation SN Ia may also arise from AM CVn binaries upon the first He flash, particularly if the donor is semi-degenerate, which is suggested by observations of AM CVn donors \citep[][]{vanRoestel2022}. Using $\mesa$ binary models \citep[][]{MESAI,MESAII,MESAIII,MESAIV,MESAV,MESAVI} that model the first dynamical He flash on the accretor, \cite{Wong2023} showed that this channel can produce hypervelocity stars like D6-2 with velocities as low as $\approx 1$,000~$\kms$. 

Recent works have modeled the stellar evolution of the surviving, previously degenerate donor with {\mesa}. 
\cite{Bhat2025} modeled the cooling of $0.5-1.1\,\msun$ inflated CO WD donors with an entropy profile based on an $\tt{AREPO}$ hydrodynamical simulation \citep[][]{Pakmor2022}. They showed that the donors contract back to $\lesssim 0.01\,\rsun$ within $\sim 10^{4}$~yrs, which is in tension with the hypervelocity stars' inferred flight-times from the disk $\sim 10^{4} - 10^{6}$~yrs. Using $\tt{AREPO}$ simulations, \cite{Glanz2024} showed that mergers between hybrid He-CO WDs can produce a $\approx 0.5\,\msun$ remnant moving at $\approx 2$,000~$\kms$, whose cooling evolution can explain some of the hottest stars. 
On the lower mass end, \cite{Bauer2019} used {\athena} \citep[][]{Stone2020} hydrodynamic simulations to model semi-degenerate He stars with masses $0.24~\&~0.35\,\msun$. \cite{Wong2024} later studied a semi-degenerate, $0.13\,\msun$ He WD, whose post-explosion evolution agrees reasonably with D6-2, although slightly more luminous at an age of $\approx 10^{5}$~yrs. 
Recently, \cite{Shen2025} modeled the Kelvin-Helmholtz contraction of He and CO stars, showing that the hypervelocity stars can be explained by WD donors with final masses $\lesssim 0.5\,\msun$, some of which may have undergone significant mass loss. 

\bigskip

Our work expands on \cite{Wong2024} and explores the parameter space of D6 stars arising from He WD donors. We show that a lower-mass WD donor suffers stronger mass loss and impact from the SN ejecta. 
Our paper is structured as follows. 
In Section \ref{sec:numerics}, we describe the numerical setup of our {\athena} simulations. 
In Section \ref{sec:theory}, we show that for Roche lobe-filling donors, a lower-mass and larger-radius donor is expected to experience a stronger impact by the SN ejecta. 
This is corroborated by the {\athena} simulation results that we show in Sections \ref{sec:model 0.08 3.9} \& \ref{sec:mass loss}. We model the post-explosion evolution of the donor with {\mesa} and present our results in Section \ref{sec:mesa}. 
We conclude in Section \ref{sec:conclusion}. 

Our {\athena} and {\mesa} input and output files and simulation movies are available in Zenodo (\href{https://doi.org/10.5281/zenodo.15493189}{doi.org/10.5281/zenodo.15493189}).

\section{Numerical Setup}
\label{sec:numerics}

Our {\athena} hydrodynamical simulations are based on \cite{Wong2024}. Here we summarize the setup and describe the major differences. 
We use {\athena} to solve the inviscid hydrodynamic equations in 3D Cartesian coordinates. 
We employ second-order time integration \citep[][]{vanLeer1979} and second-order spatial reconstruction (piecewise linear). 
We use the multigrid method for self-gravity \citep[][]{Tomida2023}. 
We adopt isolated boundary conditions, which consider only mass inside the box. This is solved via multipole expansion up to quadrapole order. 

Unlike \cite{Wong2024} which used an ideal equation-of-state (EOS) with adiabatic index $\Gamma=5/3$, here we use the general EOS module in {\athena} as described in \cite{Coleman2020}. 
Only by incorporating a true EOS can we correctly model any transition in electron degeneracy associated with the shock traversal in the donor, and the transition to radiation pressure-dominated in the shocked SN ejecta \citep[e.g.,][]{Prust2025}. 
To speed up runtime, we use a tabulated EOS based on {\mesa}, which includes the effects of radiation, degeneracy and Coulomb corrections. We describe the EOS table in detail in Appendix \ref{sec:EOS_table}. 
In our simulations, we apply pressure and energy floors corresponding locally to $\log_{10} (T/\mathrm{K}) = 5$, to avoid encountering invalid regions of the EOS \citep[see also][]{Feldman2024}. 

\subsection{Stellar models}

Our low-mass He WD models have initial masses $\iniMstar=0.08,~0.10~\&~0.126\,\msun$ and central specific entropies $\iniSc = 3.9~\&~2.7\,\kB \NA$, where $\kB$ is Boltzmann constant and $\NA$ is Avogadro's number. 
The $\iniSc = 3.9\,\kB\NA$ donors are partially degenerate whereas the $\iniSc = 2.7\,\kB\NA$ donors follow a mass-radius relation close to the fully degenerate solution. 
This range of entropy values is inspired by observations which show that AM CVn donors can be semidegenerate \citep[e.g.,][]{vanRoestel2022,Green2025}. 
We also include He WDs with $\iniMstar = 0.15 - 0.45\,\msun$. In order to avoid numerical difficulties (see Appendix), we adopt initial central temperatures of $T^{i}_{\rm c} = 2 \times 10^{7}$~K for these, corresponding to cooling ages $\approx 10^{8}$~yrs. 
Our model name convection uses $\iniMstar$ and $\iniSc$, e.g., $\verb|0.126_3.9|$. 
These initial conditions are right after any binary interaction, which we do not model. 

We adopt the initial central density and pressure of the donor as the corresponding simulation units $\inirhoc$ and $\iniPc$. The energy unit follows as $\iniPc/\inirhoc$, and velocity as $v_{0}= \sqrt{\iniPc/\inirhoc}$. Following \cite{Bauer2019} \& \cite{Wong2024}, we choose $t_{0} = \sqrt{5/(32 G \inirhoc )} $ as our time unit. The length unit is then $x_{0} = t_{0}/v_{0}$. We provide simulation parameters in Table \ref{tab:models}.

\begin{splitdeluxetable*}{ccccccccccccBc|cccccccccc}
\tablenum{1}
\label{tab:models}
\tablecaption{Helium WD Donor models and \athena\, results}
\tablehead{
\colhead{$\iniMstar$} & 
\colhead{$\iniSc$} &
\colhead{$R^{i}_{\star}$} & 
\colhead{$a$} & 
\colhead{$\Porb$} & 
\colhead{$v_{\rm orb,\star}$} &
\colhead{$\inirhoc$} &
\colhead{$\iniPc$} &
\colhead{$x_{0}$} &
\colhead{$t_{0}$} &
\colhead{$\Pej/\iniPc$} &
\colhead{$\Pej/\iniPavg$} &
\colhead{$(x,y,z)$} &
\colhead{$\Delta \mstar$} & 
\colhead{$ M^{f}_{\rm ej}$} & 
\colhead{$\rho^{f}_{\rm c}$} & 
\colhead{$\Delta s_{\rm c}$} &
\colhead{$v_{\rm{kick},x}$} & 
\colhead{$v_{\rm{kick},y}$} & 
\colhead{$v^{f}_{\star}$} &
\colhead{$v_{\rm mass~loss}$} &
\colhead{$P^{f}_{\rm rot}$} & 
\colhead{$v^{f}_{\rm rot}$}
\\
\colhead{[$\msun$]} &
\colhead{[$\kB \NA$]} &
\colhead{[$\rsun$]} &
\colhead{[$\rsun$]} &
\colhead{[min]} &
\colhead{[$\kms$]} &
\colhead{[$\gcc$]} &
\colhead{[dyne cm$^{-2}$]} &
\colhead{[$\rsun$]} &
\colhead{[s]} &
\colhead{} &
\colhead{} &
\colhead{[$x_{0}$]} &
\colhead{[$\msun$]} & 
\colhead{[$\msun$]} & 
\colhead{[$\inirhoc$]} & 
\colhead{[$\kB \NA$]} &
\colhead{[$\kms$]} & 
\colhead{[$\kms$]} & 
\colhead{[$\kms$]} &
\colhead{[$\kms$]} &
\colhead{[hr]} & 
\colhead{[$\kms$]}
}
\tablewidth{100pt}
\startdata
0.08 &   3.9 &  0.0438 &   0.226 &    17.3 &    883 & $8.94 \times 10^{3}$ & $1.66 \times 10^{19}$ &   0.0113 &   18.3 &     4.4 &      54 & $(-25,95)$, $(-60,60) \times 2$  &   0.058 &              - &   0.0028 &       0.94 &        261 &       -4.9 &        916 &       1000 &         63 &        4 \\
0.10 &   3.9 &  0.0421 &   0.204 &    14.6 &    922 & $1.34 \times 10^{4}$ & $3.28 \times 10^{19}$ &   0.0106 &   14.9 &     3.1 &      40 & $(-15,85)$, $(-50,50) \times 2$ &   0.039 &                - &    0.037 &       0.58 &        311 &       -3.9 &        970 &       1200 &         24 &         11 \\
0.126 &   3.9 &  0.0410 &   0.185 &    12.6 &    955 & $2.02 \times 10^{4}$ & $6.46 \times 10^{19}$ &  0.00988 &   12.2 &     2.1 &      29 & $(-15,85)$, $(-50,50) \times 2$ &   0.026 & $2 \times 10^{-7}$ &     0.13 &       0.46 &        300 &       -5.0 &        997 &       1400 &         17 &         17 \\
0.08 &   2.7 &  0.0307 &   0.158 &    10.1 &   1055 & $2.61 \times 10^{4}$ & $7.11 \times 10^{19}$ &  0.00802 &   10.7 &       3.0 &      37 & $(-15,85)$, $(-50,50) \times 2$ &   0.025 &                - &     0.06 &       1.12 &        342 &       -5.8 &       1104 &       1400 &         20 &         12 \\
0.10 &   2.7 &  0.0296 &   0.143 &     8.6 &   1100 & $4.03 \times 10^{4}$ & $1.47 \times 10^{20}$ &  0.00745 &    8.6 &       2.0 &      27 & $(-15,85)$, $(-50,50) \times 2$ &  0.014 & $1 \times 10^{-5}$ &     0.17 &       0.94 &        321 &       -7.6 &       1139 &       1500 &         14 &         17 \\
0.126 &   2.7 &  0.0281 &   0.127 &     7.1 &   1154 & $6.37 \times 10^{4}$ & $3.13 \times 10^{20}$ &  0.00689 &    6.8 &     1.3 &      20 & $(-15,85)$, $(-50,50) \times 2$ &   0.011 & $3 \times 10^{-5}$ &     0.32 &       0.75 &        295 &       -9.3 &       1182 &       1900 &         31 &         13 \\
0.15 &   3.1 &  0.0319 &   0.137 &     7.9 &   1098 & $7.05 \times 10^{4}$ & $3.90 \times 10^{20}$ &  0.00695 &    6.5 &    0.84 &      17 & $(-30,60)$, $(-45,45) \times 2$ &    0.010 & $4 \times 10^{-5}$ &     0.45 &       0.38 &        173 &      -12.2 &       1100 &       1800 &          8.0 &         26 \\
 0.20 &   2.8 &  0.0260 &   0.103 &     5.1 &   1240 & $1.58 \times 10^{5}$ & $1.41 \times 10^{21}$ &   0.0059 &    4.3 &    0.55 &     9.9 & $(-30,50)$, $(-40,40) \times 2$ &  0.0075 & $8 \times 10^{-5}$ &      0.60 &       0.37 &        123 &      -18.8 &       1227 &       2300 &       0.78 &         94 \\
 0.25 &   2.6 &  0.0226 &   0.084 &     3.7 &   1344 & $2.82 \times 10^{5}$ & $3.56 \times 10^{21}$ &  0.00525 &    3.3 &     0.40 &     6.7 & $(-30,50)$, $(-40,40) \times 2$ &  0.0063 & $2 \times 10^{-4}$ &     0.69 &       0.35 &         81 &      -26.3 &       1320 &       2900 &        0.30 &        164 \\
 0.30 &   2.5 &  0.0204 &   0.073 &     2.9 &   1420 & $4.50 \times 10^{5}$ & $7.50 \times 10^{21}$ &  0.00477 &    2.6 &     0.30 &     4.9 & $(-30,50)$, $(-40,40) \times 2$ &  0.0056 & $5 \times 10^{-4}$ &     0.77 &       0.33 &         42 &      -33.3 &       1388 &       3400 &       0.16 &        233 \\
 0.35 &   2.4 &  0.0185 &   0.063 &     2.3 &   1493 & $6.73 \times 10^{5}$ & $1.42 \times 10^{22}$ &  0.00439 &    2.1 &    0.24 &     3.8 & $(-30,50)$, $(-40,40) \times 2$ &  0.0053 & $9 \times 10^{-4}$ &     0.79 &       0.33 &         11 &      -40.5 &       1452 &       3900 &       0.11 &        302 \\
 0.40 &   2.3 &  0.0171 &   0.056 &     1.9 &   1554 & $9.64 \times 10^{5}$ & $2.48 \times 10^{22}$ &  0.00406 &    1.8 &    0.19 &       3.0 & $(-30,50)$, $(-40,40) \times 2$ &  0.0053 & $1 \times 10^{-3}$ &     0.81 &       0.32 &        -17 &      -47.9 &       1507 &       4300 &      0.075 &        375 \\
 0.45 &   2.2 &  0.0159 &   0.051 &     1.6 &   1608 & $1.34 \times 10^{6}$ & $4.14 \times 10^{22}$ &  0.00377 &    1.5 &    0.16 &     2.4 & $(-30,50)$, $(-40,40) \times 2$ &  0.0051 & $2 \times 10^{-3}$ &     0.83 &       0.30 &        -50 &      -54.5 &       1554 &       4600 &      0.056 &        451 \\
\enddata
\tablecomments{\textit{Top: initial conditions.} $\iniMstar$, $s_{\rm c}^{i}$ and $R^{i}_{\star}$ are the initial donor mass, central specific entropy and radius, $a$ and $\Porb$ are the binary separation and orbital period, $v_{\rm orb,\star}$ is the orbital velocity of the donor, $\rho^{i}_{\rm c}$ and $P^{i}_{\rm c}$ are the initial donor central density and pressure, $x_{0}$ and $t_{0}$ are the length and time units of the simulation, $\Pej/\iniPc$ is the ratio between SN ejecta ram pressure and initial donor central pressure, $\Pej/\iniPavg$ is the ratio between ejecta ram pressure and initial donor volume-averaged pressure, and $(x,y,z)$ is the domain of our {\athena} simulations. 
\textit{Bottom: simulation results.}
$\Delta \mstar$ is the donor mass loss, $\Mej^{f}$ is the bound ejecta mass (we do not show values $<10^{-10}\,\msun$), $\rho^{f}_{\rm c}$ is the final donor central density, $\Delta s_{\rm c}$ is the jump in donor central specific entropy, $v_{\rm kick,x}$ \& $v_{\rm kick,y}$ is the kick velocity received by the donor in the $x$- \& $y$-directions, $v^{f}_{\star}$ is the final donor velocity after accounting for the kick and orbital velocities, $v_{\rm mass~loss}$ is the median velocity of the donor mass loss, and $P^{f}_{\rm rot}$ and $v^{f}_{\rm rot}$ are the estimated surface rotation period and velocity at $10^{5}$~yrs. } 
\end{splitdeluxetable*}

\subsection{Modeling ejecta-companion interaction}

After mapping the donor onto the grid, we allow relaxation for 20 time units, with velocity damping as detailed in \cite{Wong2024}. We further damp spurious velocities in low density regions, so as to prevent time steps being limited. For simplicity, we do not account for accretor gravity and centrifugal acceleration in the co-rotating frame. A detailed exploration of their effects is presented in \cite{Wong2024}, who showed that a companion distorted by the Roche potential can have slightly lower mass loss, especially with smaller mass ratios.

The SN ejecta is injected into the grid after donor relaxation following \cite{Wong2024}. 
We place the explosion center along the $x$-axis at a binary separation $a$ such that the donor is Roche lobe-filling given a $1\,\msun$ accretor. 
We adopt ejecta mass $\mwd = 1 \, \msun$, total kinetic energy $\Eej = 1.2 \times 10^{51}$~erg, and a Gaussian density profile assuming homology, which \cite{Wong2024} showed agreed well with previous detonation models \citep[e.g.,][]{Seitenzahl2013,Shen2018_subMch}. 
We set the ejecta pressure profile as 
$P(v,t) = \langle P/\rho^{5/3} \rangle  (\rho(v,t))^{5/3}$, where $\langle P / \rho^{5/3} \rangle = 0.59 \times 10^{14}$ in cgs units is a spatially averaged value from the \cite{Shen2018_subMch} models. 
We exclude ejecta faster than 20,000~km/s. 
Our simulations consider a frame centered on the donor at the time of explosion, and we account for the orbital motion by adding a negative $y$-velocity in the ejecta, assuming that the binary is instantly unbound (however, see \citealt{Braudo2024}). 

We follow the hydrodynamical evolution of the donor until its oscillations damp out and it largely returns to hydrostatic equilibrium. Due to the interaction with SN ejecta, the donor gains a kick velocity (mostly in $x$). Once the donor moves to the center of the simulation box, we re-center our reference frame and subtract the center-of-mass velocity, keeping the donor in the box.


\section{Ejecta-companion Interaction}
\label{sec:athena_results}

\subsection{Theoretical expectations}
\label{sec:theory}

Here we review the expected impact of the SN ejecta on the donor, first presented in \cite{Bauer2019}. 
The ram pressure of the SN ejecta at the donor location, is $\Pej \approx \Eej/(4 \pi a^{3}/3)$. 
Assuming an $n=1.5$ polytrope, the donor central pressure is $P_{\rm c} = 0.77 G M_{\star}^{2}/R_{\star}^{4}$. 
When Roche lobe-filling, the ratio of the donor radius to the binary separation is given by $R_{\star} / a \approx 0.462 (q/(1+q))^{1/3}$ \citep[][]{Paczynski1971}, where the mass ratio is given by $q \equiv \mstar / \Mej$ and $\Mej$ is the ejecta (accretor) mass. 
This yields
\begin{align}
\label{eqn:ratio}
\frac{P_{\rm ej}}{ P_{\rm c} } \approx \, & 4 (1+q)^{-1} \left( \frac{ \rstar }{ 0.05 \, \rsun } \right) \left( \frac{ \mstar }{ 0.1 \, \msun } \right)^{-1} \nonumber\\
&\times \left( \frac{E_{\rm ej}}{10^{51}\,\mathrm{erg}} \right) \left( \frac{ \Mej }{ 1 \, \msun } \right)^{-1} .
\end{align}
This scaling shows that Roche lobe-filling donors in wider binaries experience a stronger impact from the SN ejecta. Although one might intuitively expect the opposite because the explosion energy $E_{\rm ej}$ is diluted by the volume $a^{3}$, we obtain this result because the donor is \textit{Roche lobe-filling}. Its gravitational binding energy $E_{\rm g} \sim - G \mstar^{2} / \rstar$ has a smaller magnitude for a larger radius (i.e. more weakly bound), and is similarly spread over a volume $\rstar^{3} \propto a^{3}$. 
Furthermore, equation \ref{eqn:ratio} shows that lower-mass Roche lobe-filling donors suffer a stronger impact. For non-relativistic, fully degenerate donors with mass-radius relation $\rstar \propto \mstar^{-1/3}$, we expect that $P_{\rm ej} / P_{\rm c} \propto \mstar^{-4/3} $, and thus a much stronger impact for lower-mass WD donors. 
We provide the values of $\Pej/\iniPc$ \& $\Pej/\iniPavg$, where $\iniPc$ and $\iniPavg$ are the initial central and volume-averaged pressures of the initial stellar model, in Table \ref{tab:models}. 
In the following, we show that indeed, as we lower the mass of the He WD donor, it undergoes stronger mass loss and impact by the ejecta. 

\subsection{Model with strongest impact}
\label{sec:model 0.08 3.9}


\begin{figure*}[t!]
\centering
\fig{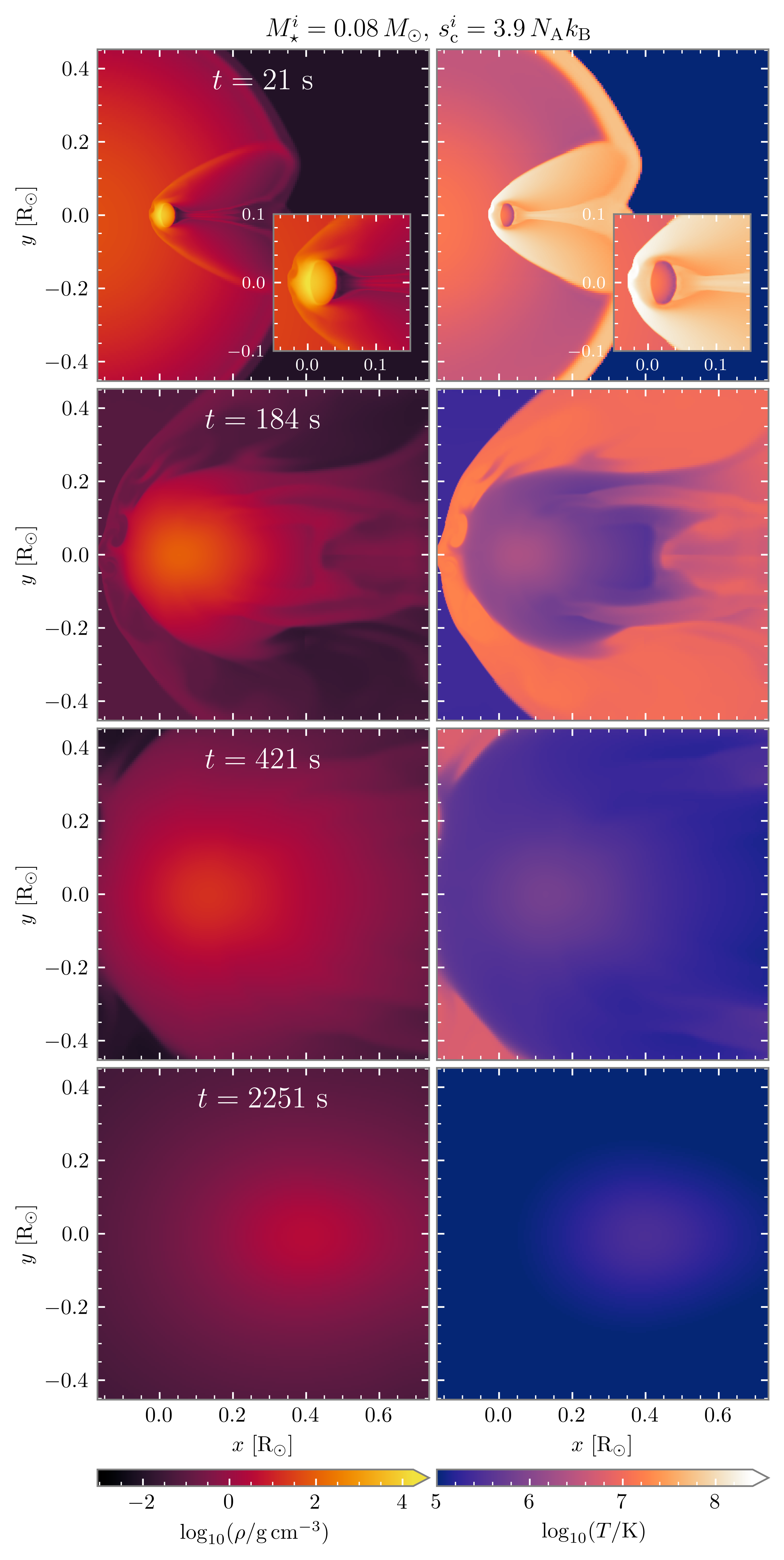}{ 0.6 \textwidth }{}
\caption{ 
Time evolution of the density (left) and temperature (right) at the mid-plane for the $\iniMstar=0.08\,\msun$, $\iniSc=3.9\,\NA\kB$ simulation. Timestamps in the left column give the time since explosion. The insets in the top row show zoomed-in views around the donor. 
In the first row, the outer edge of the ejecta consists of swept-up, shocked floor material which is unimportant to the ejecta-companion interaction. 
\label{fig:TimeSequence}}
\end{figure*}


We illustrate the interaction between ejecta and donor using model \verb|0.08_3.9|, which has the lowest mass and largest radius and hence reveals the strongest impact on the donor. 
Figure \ref{fig:TimeSequence} shows the time evolution of the mid-plane density and temperature at various times since explosion. 
As the ejecta encounters the donor, a bow shock is formed with half-opening angle $\approx \arctan(2\rstar/a) \approx 20^{\circ}$ \citep[][]{Kasen2010,Prust2025} and post-shock temperatures $\approx \rm{few} \times 10^{8}$~K in the star and ejecta (top row). A low-density wake forms behind the donor, which introduces asymmetry in the remnant that persists for thousands of years \citep[][]{Prust2025}. 
As seen more clearly in the inset, a forward shock is sent into the donor and travels through its interior. Shock breakout from the downstream side of the donor unbinds some material, but most of the unbound donor material originates from the upstream hemisphere \citep[][]{Wong2024}. The mass loss is highly anisotropic -- most of it is contained in the wake behind the donor, peaking at an angle $\approx 20 - 40^{\circ}$ from the binary axis \citep[][]{Marietta2000,Wong2024}. 

Entropy deposition by the shock traversal causes the donor to expand (second row), while the bow shock widens and its standoff distance from the donor center increases \citep[][]{Prust2024}. 
The shocked donor continues expanding and its central density reaches a minimum of $\approx 3 \,\gcc$ at $t \approx 2$,200~s (bottom row). 
Only $\approx 0.02\,\msun$ remains bound. 
Due to expansion, the outermost, bound parts of the donor reach the temperature floor of $10^{5}$~K , but the central parts remain hotter and well-resolved. 

The donor remnant undergoes periodic radial pulsations as it recovers hydrostatic equilibrium. The pulsations send out shock waves into the still infalling outer bound material. Meanwhile, the center undergoes adiabatic compression and expansion, and its specific entropy is roughly constant at the post-shock value. Eventually the pulsations subside and the donor remnant achieves hydrostatic equilibrium. 
We report the total \textit{unbound} mass that leaves the box as $\Delta \mstar$ in Table \ref{tab:models}. 


\begin{figure*}
\centering
\fig{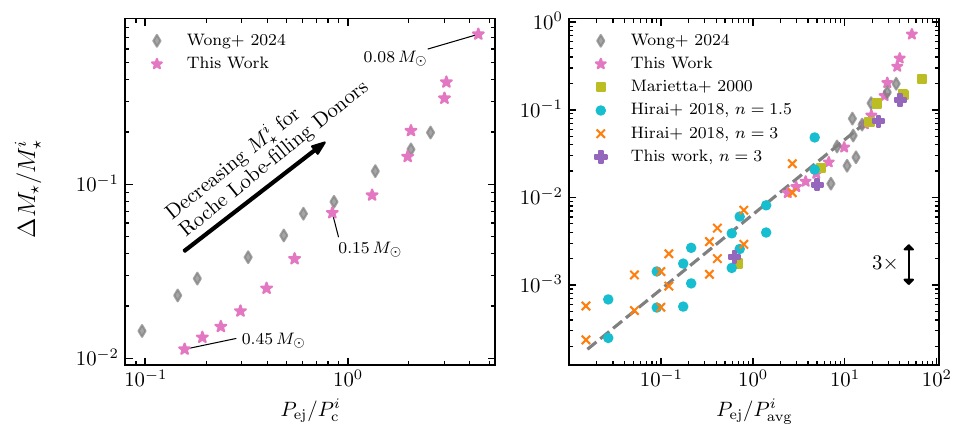}{ \textwidth }{}
\caption{ 
Fractional mass loss of the donor, $\Delta \mstar/ \iniMstar$, as a function of the ratio between ejecta ram pressure and initial donor central pressure, $\Pej/P^{i}_{\rm c}$ (left), and ratio between ejecta ram pressure and initial donor volume-averaged pressure, $\Pej/P^{i}_{\rm avg}$  (right). 
The left panel includes models from this work (pink stars) and \cite{Wong2024} (grey diamonds; $\Gamma = 5/3$). Roche lobe-filling donors with lower $\iniMstar$ are towards the right. In the right panel, we include the main sequence models from \cite{Marietta2000} (their HCV model; yellow squares), the $n=1.5$ (blue circles) and $n=3$ (orange crosses) polytrope models from \cite{Hirai2018}, our $n=3$ polytrope models (purple plus signs), and our fitted power-law relation (grey dashed line; equation \ref{eqn:mass_loss}) in the right panel. The scatter in the data spans a factor of $\approx 3$. 
\label{fig:delta_M}}
\end{figure*}


All simulations show qualitatively similar evolution as described above, as do 
other previous works on ejecta-companion interaction \citep[e.g.,][]{Marietta2000,Pakmor2008,Pan2010,Liu2012,Hirai2018,Bauer2019,Rau2022,Wong2024,Pan2025}. 
The results of our simulations are summarized in Table \ref{tab:models}. 

As the donor becomes increasingly impacted by the ejecta and we approach the limit of complete disintegration, where the donor becomes completely unbound \citep[see also][]{Pan2025}, we find that more and more of the mass leaving the box is still in the bound state. In the most extreme case, model \verb|0.08_3.9|, the bound mass loss is $\approx 20\%$ of the total mass loss, so there is a growing uncertainty on the mass loss as we approach the disintegration limit. 
The amount of additional mass loss due to this effect is similar to that due to coarser numerical resolution of the hydrodynamical simulations, where we find a small impact on the shocked donor's subsequent stellar evolution, as seen from similar temperature-density profiles of the shocked donor and similar evolutionary tracks on the Hertzsprung-Russell diagram. 
In addition, \cite{Wong2024} vary the box size and find small variations in the central density of the shocked donor (see their Appendix A). 
Long after the ejecta leaves the box, we find that the amplitude of the total energy in the box (which is then dominated by the surviving donor) decreases by $\lesssim$ a few percent due to the bound mass loss. 
Overall, the donor material that leaves the box is still dominated by unbound material ($\gtrsim 80\%$). 
Therefore, the small amount of bound mass loss is likely insignificant to our models, especially with the higher-mass models $\iniMstar \gtrsim 0.15\,\msun$.


\subsection{Shock jump and mass loss}
\label{sec:mass loss}

With a higher $\Pej/\iniPc$ ratio, the donor experiences a stronger impact by the ejecta. 
The jump in central density, as the shock wave passes, varies between 1.5 and 2.3, and the pressure jump between 2 and 6, reflecting upstream Mach numbers $\approx 1.3 - 2.2$ assuming $\Gamma = 5/3$. Both increase with $\Pej/\iniPc$, showing a stronger shock and hence impact on the donor.

The fractional mass loss of the donor $\Delta \mstar / \iniMstar$ increases with $\Pej/\iniPc$, as shown in the left panel of Figure \ref{fig:delta_M}. 
We include models from \cite{Wong2024} where a range of ejecta kinetic energies $\Eke = 0.5 - 1.5 \times 10^{51}$~erg is taken. The donor models include $0.24$ \& $0.33\,\msun$ He stars, and the same He WD as our model \verb|0.126_3.9|. 
Some scatter in the $\Delta \mstar / \iniMstar - \Pej/\iniPavg$ relation exists, because of different modeling assumptions (see Section \ref{sec:numerics}) and donor structures. 
The donors can be approximated as polytropes with indices $n\approx1.5-2.3$, close to the fully degenerate, non-relativistic solution $n=1.5$ and varying due to various cooling times. 
Moreover, having undergone core He burning, the He star models are chemically inhomogeneous. 
These models show a tight relation between $\Delta \mstar/\iniMstar$ and $\Pej/\iniPc$, though extending this to other companion types, e.g. a main sequence star, introduces some scatter.

To facilitate comparison to models where the radial density structure of the stars vary, we use instead $\Pej/\iniPavg$ as our diagnostic (right panel, Figure \ref{fig:delta_M}). For this comparison, we also include the main sequence and subgiant models from \cite{Marietta2000} (their HCV and HCVL models), and the $n=1.5~\&~3$ polytrope models from \cite{Hirai2018}. We additionally performed four simulations with an $n=3$ polytrope with mass $1\,\msun$ and radius $1\,\rsun$, and varying separations $a=2.5,\,3,\,5\,\&10\,\rsun$. For these runs we adopt a $\Gamma=5/3$ EOS and keep the ejecta properties the same as our fiducial runs. The model parameters are only slightly different than the HCV series of \cite{Marietta2000}, and we find good agreement with their reported amount of mass loss. 

Despite the large differences in companion structure, ejecta profile, and modeling assumptions across these works, $\Delta \mstar / \iniMstar$ follows a relatively tight relation with $\Pej/\iniPavg$. For $\Pej/\iniPavg \lesssim 20 $, we find a power-law relation
\begin{equation}
\label{eqn:mass_loss}
    \frac{\Delta \mstar}{\iniMstar} = 5.8\times10^{-3} \left( \frac{\Pej}{\iniPavg} \right)^{0.78} .
\end{equation}
This illustrates that a companion with a larger radius and lower mass is more strongly impacted by SN ejecta. 
Some subtleties arise from different companion structures. At the most extreme end of $\Pej/\iniPavg$, our He WD models show more mass loss than our $n=3$ polytrope models and the main sequence and subgiant models of \cite{Marietta2000}, by factor of a few. This can be attributed to the centrally concentrated structure of the latter: due to the steeper density gradient, the shock in the donor weakens more as it reaches the stellar interior, and is thus unable to unbind a substantial fraction of the donor.

For Roche lobe-filling CO WD donors, equation \ref{eqn:mass_loss} predicts little mass loss, and we can compare to \citealt{Tanikawa2019} where $\Mej = 1\,\msun$, $\Eej \approx 1.1 \times 10^{51}$~erg. For their $0.6\,\msun$ CO WD donor ($\iniPavg  \approx 4.2 \times 10^{22}$ dyne cm$^{-2}$, $a \approx 0.036 \, \rsun$)\footnote{We estimate the average pressure using the cold WD models from Frank Timmes: \href{https://cococubed.com/code_pages/coldwd.shtml}{https://cococubed.com/code\_pages/coldwd.shtml}}, we predict a mass loss of $\approx 2 \times 10^{-3} \,\msun$, in excellent agreement with their $\approx 3 \times 10^{-3}\,\msun$. For their $0.9\,\msun$ CO WD ($\iniPavg  \approx 4.1 \times 10^{23}$ dyne cm$^{-2}$, $a \approx 0.023\,\rsun$), our predicted $\approx 1 \times 10^{-3}\,\msun$ is also close to their reported $\approx 3 \times 10^{-3}\,\msun$.

We compare the mass loss in our simulations to the analytical models by \cite{Wheeler1975}. Similarly to \cite{Marietta2000}, \cite{Hirai2018} \& \cite{Rimoldi2016}, we find that the analytical models overpredict the mass loss, sometimes by an order of magnitude, and does not scale well with binary separation $a$. 

Recently, in the context of star-accretion disk collisions, \cite{Yao2025} derived a similar scaling $\Delta \mstar / \iniMstar \propto P_{\rm ram} / \iniPavg$, where $P_{\rm ram}$ is the ram pressure, by considering the fraction of stellar mass that is shocked by the collision. Our numerical study agrees with this scaling. The same scaling was also noted by \cite{Linial2023} for the ejecta-companion interaction work by \cite{Liu2015}. Moreover, our findings are consistent with findings from previous works that $\Delta M \propto \Eej$ \citep[][]{Pakmor2008,Liu2013c,Rau2022} and $\Delta M \appropto a^{-3} $ \citep[][]{Pakmor2008,Pan2010,Pan2012a,Liu2012,Liu2013a,Liu2013c,Rau2022,Pan2025}. 

Our most extreme case, model \verb|0.08_3.9|, loses $\approx 70\%$ of its initial mass. 
While it is interesting to see whether a He WD donor would completely unbound once $\Pej/\iniPavg$ reaches a sufficiently large value $\gtrsim 60$ \citep[see also][for related work on M dwarf companions]{Pan2025}, we leave this for future work, and note that tracking the hydrodynamic evolution of the significantly expanded donor requires a much larger simulation box.

\subsection{Velocity distribution of unbound donor material in ejecta}
\label{sec:velocity distribution}

A growing number of observational studies have placed tight limits on narrow emission lines in the nebular phase ($\gtrsim$100~days) resulting from unbound He-rich donor material, providing a method to distinguish among various binary progenitors \citep[e.g.,][]{Mattila2005,Lundqvist2013,Lundqvist2015,Maguire2016,Graham2017,Sand2018,Dimitriadis2019,Tucker2019,Tucker2020,Sand2021,Hosseinzadeh2022}. This motivates us to study the velocity distribution of the mass loss, by recording the flux of unbound donor material (Bernoulli parameter $< 0$) leaving the simulation box at various velocity bins. We account for the donor orbital motion when calculating the velocity distribution in an inertial frame centered on the binary center-of-mass \citep[see also][]{Wong2024}. 

\begin{figure}
\centering
\fig{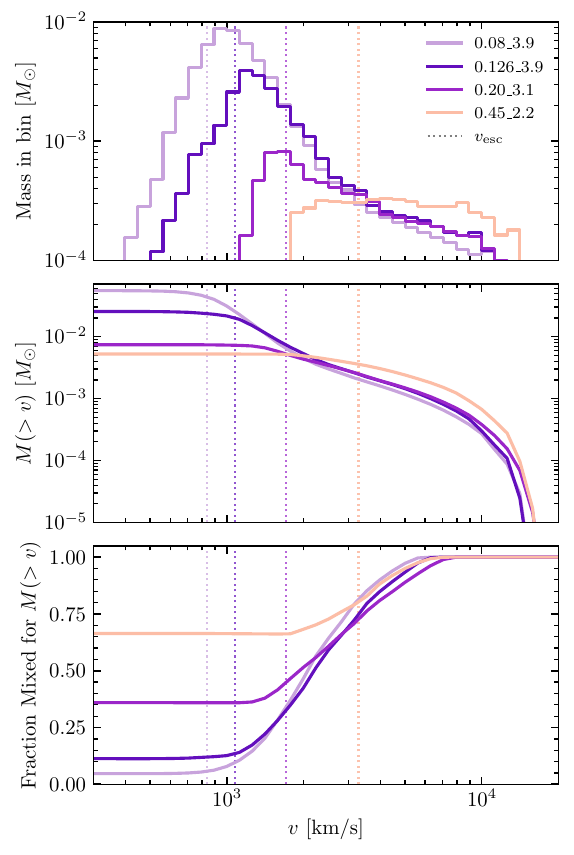}{ 0.5\textwidth }{}
\caption{ 
Velocity distribution of donor mass loss (top), total donor mass loss above a given velocity (middle), and the fraction of mixed donor mass loss above a given velocity (bottom), for selected models. Dotted lines show the escape velocity of the unperturbed donor. 
\label{fig:vel_dist}}
\end{figure}

The top panel of Figure \ref{fig:vel_dist} shows the velocity distribution of the mass loss for selected models. We provide the median velocity of the mass loss of all models in Table \ref{tab:models}. We find that most of the mass loss occurs near the escape velocity $v_{\rm esc} = \sqrt{2 G \mstar / \rstar}$ of the donor. Since higher-mass donors are more compact, the median velocity increases from $\approx$~1,000\,$\kms$ for $\iniMstar=0.08\,\msun$, to $\approx$~4,600\,$\kms$ for $\iniMstar=0.45\,\msun$, notably faster than H-rich donors \citep[$400-900\,\kms$;][]{Marietta2000,Boehner2017,McCutcheon2022}. The higher-mass models also show a wider distribution. 

We show the total mass loss above a given velocity in the middle panel of \ref{fig:vel_dist}. We find a high velocity tail, $\approx$~few~$ 10^{-4}\,\msun$, of stripped He lying above 10,000~$\kms$, roughly independent of donor model. Model \verb|0.45_0.22| has slightly more mass in the high velocity tail, due to its higher $v_{\rm esc}$. The $\approx 10^{-4}\,\msun$ of high-velocity tail has also been seen in simulations with H-rich companions \citep[][]{Marietta2000,Boehner2017}. 

Since mass loss of higher $\iniMstar$ lies at higher velocities, lost donor material is comoving with SN ejecta. To explore the degree of mixing between SN ejecta and donor material, we record the velocity distribution of donor material that is contaminated by SN ejecta, which we define as donor material where the donor mass fraction $< 0.9$. The fraction of mixed donor material above a given velocity is given in the bottom panel of Figure \ref{fig:vel_dist}. All models show mixed donor material at $v \gtrsim $~6,000~$\kms$, which amounts to $\approx 10^{-3}\, \msun$ of He. The degree of mixing decreases towards lower velocities, as donor material at lower velocities is more shielded from the ejecta. 
The least massive model, \verb|0.08_3.9|, shows a low degree of mixing overall, as most of its mass loss lies at $\approx$~1,000$\,\kms$, much slower than the bulk ejecta. On the other hand, the most massive model, \verb|0.45_2.2|, has more than half of its mass loss mixed to some degree with ejecta, as its mass loss is concentrated at higher velocities.

Only one nebular phase radiative transfer calculation has been performed for He-rich material \citep[][]{Botyanszki2018}, leading to inferred upper limits of stripped He material typically $\lesssim 10^{-3}\,\msun$ in SNe Ia and Iax \citep[e.g.,][]{Tucker2020,JacobsonGalan2019}. However, the velocity distribution $\lesssim $~1,000~$\kms$ is obtained from the H-rich main sequence companion simulation by \cite{Boehner2017}, by replacing H with He. Our work shows a faster velocity distribution where the typical velocity and mixing with SN ejecta are correlated with the amount of mass loss. The degree of mixing between ejecta and He affects the nonthermal excitation of He I \citep[][]{Lucy1991,Dessart2012}. 
Furthermore, \cite{Dessart2020_H} predicted no optical He I lines, in conflict with the prediction of strong optical He I lines by \cite{Botyanszki2018}, with the difference likely due to optical depth effects. 
These effects may affect the predicted line luminosities \citep[e.g.,][]{Dessart2012,Dessart2020_H}, and should be considered in future radiative transfer calculations for placing limits on He mass within the ejecta.


\subsection{Final donor velocity}
\label{sec:final speed}

The donor receives a kick velocity mostly in the $x$-direction in the sense of Figure \ref{fig:TimeSequence}, $v_{\rm kick,x}$, and to a smaller extent in the $y$-direction, $v_{\rm kick,y}$, since we account for the orbital motion of the ejecta. The final donor velocity is then $v^{f}_{\star} = \sqrt{v_{\rm kick,x}^{2} + (v_{\rm kick,y} + v_{\rm orb,\star})^{2}}$. We report these quantities in Table \ref{tab:models}. 

The $x$-kick velocity is given by $\approx 1/3$ of the momentum intercepted by the donor \citep[][]{Hirai2018}. For the low-mass models, we find an efficiency roughly in this range \citep[see also][]{Bauer2019,Wong2024}, with $v_{\rm kick,x} \approx 200-300~\kms$. 
Since $v_{\rm orb,\star}$ decreases with lower $\iniMstar$, the effect of the kick velocity is more significant for these donors \citep[][]{ElBadry2023}. 

The high-mass models are increasingly affected by the gravitational acceleration of the ejecta $\sim - G \Mej / a^{2}$ before it leaves the binary orbit, leading to a negative $v_{\rm kick,x}$. This is a numerical artifact since we do not properly model the unbinding of the binary. \cite{Braudo2024} found that the donor may slow down by $\approx 10\%$ due to the ejecta. 

The $y$-kick velocity is opposite to the donor orbital velocity, and is stronger with higher $\iniMstar$ due to the tighter orbit. In some cases, it leads to a slight slowdown of the donor due to the drag from ejecta \citep[][]{Bauer2019,Wong2024}. 

Since the kick velocity acts in perpendicular to the donor orbital velocity, it does not lead to a significant speedup of the donor. The ejection velocity of the hypervelocity stars is almost entirely determined by the donor orbital velocity.

\section{Post-shock Evolution in MESA}
\label{sec:mesa}

\begin{figure*}[t!]
\centering
\gridline{
\fig{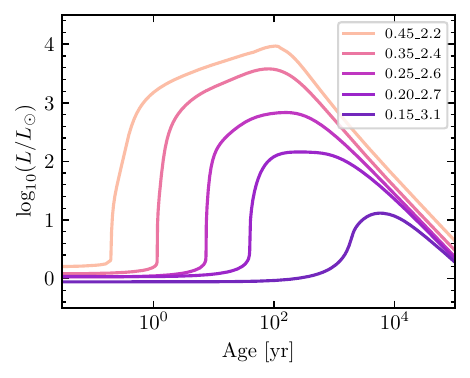}{0.475\textwidth}{}
\fig{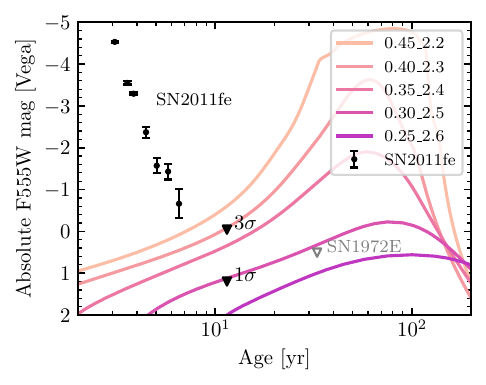}{0.475\textwidth}{}
}
\caption{ 
Time evolution of the luminosity (left) and F555W absolute magnitude for shocked donors with initial masses $\iniMstar=0.15-0.45\,\msun$. For comparison, we show the light curve of SN2011fe from \cite{Tucker2024}, including the 1 \& 3 $\sigma$ (filled triangles) nondetection limits at 11.5 yrs post-explosion, and we adopt distance modulus of $\mu = 29.03$~mag. The nondetection limit of SN1972E \citep[$\mu=27.49$~mag;][]{Do2021} at 33~yrs after explosion is shown as a grey open triangle.
\label{fig:time_evol}}
\end{figure*}

\begin{figure*}
\centering
\gridline{
\fig{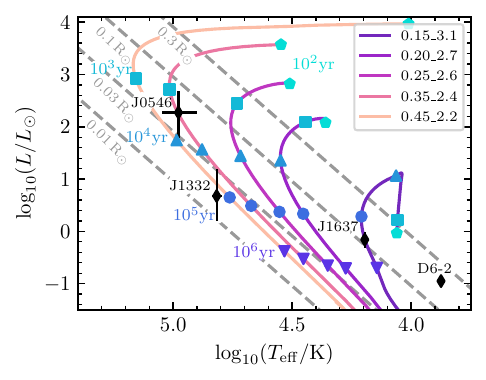}{0.475\textwidth}{}
\fig{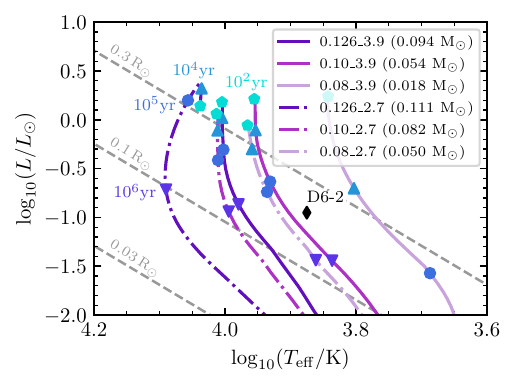}{0.5\textwidth}{}
}
\caption{ 
Evolution on the Hertzsprung-Russell diagram of the shocked donors with initial masses $\iniMstar=0.15-0.45\,\msun$ (left) and $0.08-0.126\,\msun$ (right), starting from $10^{2}$~yrs. Models on the right have initial central specific entropies $\iniSc = 3.9$ (solid lines) \& $2.7\,\kB\NA$ (dashed lines), and we show their final masses in the legend. For each model we label ages $10^{2}$ (pentagon),$10^{3}$ (square), $10^{4}$ (upright triangle), $10^{5}$ (circle) \& $10^{6}$~yrs (downright triangle). We also show lines of constant radius (dashed) and several hypervelocity stars (black diamonds). 
\label{fig:HR}}
\end{figure*}

After the hydrodynamical interaction between ejecta and donor, we model the shocked donor's subsequent evolution using the stellar evolution code {\mesa} version 24.03.1 \citep[][]{MESAI,MESAII,MESAIII,MESAIV,MESAV,MESAVI}. We first relax the mass of the stellar model to the remaining bound mass in the last snapshot of the {\athena} simulation, and then relax the $\rho-T$ profile to the corresponding spherical averages in {\athena}, as initial conditions. Our {\mesa} models include nuclear reactions, but we do not find nuclear burning sufficient enough to modify the donor's post-explosion evolution. If the earlier shock interaction leads to significant nuclear energy release, the donor's post-explosion thermal profile may change, and burning on a longer timescale may occur. 

The post-explosion donor entropy profiles are similar to \cite{Bauer2019}, \cite{Wong2024} \& \cite{Bhat2025}: the jump in specific entropy increases towards the surface \citep[see also][]{Lu2025}, reflecting the stronger strength of the shock sent by the SN impact as it traverses through the outermost, low-density regions. The entropy of the outermost bound material is further increased due to the shocks created by the oscillations of the donor. As a result, the region of peak heating, and hence the temperature peak, lies in the envelope of the donor. Due to the entropy profile, the stellar interior is convectively stable. 

As heat diffuses outwards from the deeper layers, the high-mass ($\iniMstar \geqslant 0.15 \, \msun$) models brighten and expand significantly, as shown by the left panel of Figure \ref{fig:time_evol}. 
As shown by \cite{Zhang2019}, the brightening timescale is set by the local minimum in the thermal timescale in the high-entropy envelope, which decreases with stellar mass. 
As a result, higher-mass models brighten on a shorter timescale, $\sim 1$~yr for $\iniMstar \gtrsim 0.35\,\msun$, and reach higher peak luminosities closer to the Eddington limit. 
During this phase, wind mass loss may occur due to the high luminosity, in addition to energy deposition by radioactive decay of accreted SN ejecta \citep[][]{Bhat2025,Glanz2024}, which we do not model. The low-mass ($\iniMstar \lesssim 0.126\,\msun$) models do not show this brightening. 

Figure \ref{fig:HR} shows the model tracks on the Hertzsprung-Russell (HR) diagram, starting from 100~yrs after explosion. 
Models with $\iniMstar\gtrsim0.15\,\msun$ lose little mass $\lesssim 0.01\,\msun$ due to SN impact, whereas the lower-mass models lose a significant portion of their masses, so we also show their final masses in the legend. 

The shocked donor remains luminous on a timescale set by the local heat diffusion timescale at the location of peak heating \citep[][]{Zhang2019,Bauer2019}, ranging from $\sim 10^{2}$~yr for model \verb|0.45_2.2|, to $\sim 10^{4}$~yrs for model \verb|0.15_3.1|. 

Then shocked donors undergo Kelvin-Helmholtz contraction and eventually return to the WD cooling track. Models with $\iniMstar\gtrsim0.3\,\msun$ return to the WD cooling track within $\approx 10^{3}$~yrs, in agreement with \cite{Bhat2025}. Donors with lower post-explosion masses return to the WD cooling track on longer timescales, taking $\gtrsim 10^{5}$~yrs. 

Our models are similar to the He-rich cooling models of \cite{Shen2025}, with some differences especially in the initial evolution. This is because \cite{Shen2025} assume fully convective initial models, whereas our shocked donors have radiative interiors, since the post-shock entropy profile increases towards the surface due to stronger shock-heating \citep[][]{Bauer2019,Wong2024,Bhat2025,Lu2025}. 

Our {\mesa} models are non-rotating, but we can estimate the donor surface rotation assuming uniform rotation of the tidally locked pre-explosion donor, and that the specific angular momentum at each mass coordinate is conserved. 
For an initial radius $R_{0}$ of the mass coordinate that corresponds to the post-explosion surface layer with radius $R^{f}_{\star}$, the post-explosion surface rotation period is then $P^{f}_{\rm rot} = \Porb (R^{f}_{\star}/R_{0})^{2}$. The estimated rotation periods and velocities at $10^{5}$~yrs are given in Table \ref{tab:models}. This simple estimate neglects angular momentum transport during the subsequent evolution.


\subsection{Detectability in nearby SNe Ia and remnants}
\label{sec:detectability}

The short brightening timescale of the high-mass ($\gtrsim 0.3\,\msun$) models allows for possible detection in nearby SNe Ia within a few years of explosion. For instance, SN2011fe, one of the most well-studied SN Ia occurring in the nearby galaxy M101, has deep imaging by the Hubble Space Telescope (HST) spanning 11.5 years, which ruled out most non-degenerate donors \citep[][]{Tucker2024}. To facilitate comparison to observations, we compute synthetic HST F555W (Vega) magnitude for our models using {\tt STSYNPHOT} \citep[][]{stsynphot}, assuming a blackbody and applying no extinction. The resulting absolute magnitudes are shown in the right panel of Figure \ref{fig:time_evol}. We show the HST F555W observations published by \cite{Shappee2017} \& \cite{Tucker2022} on days $\approx$~1,100$-$2,400, as well as the $1\sigma$ \& $3\sigma$ non-detection limits of 29.0 \& 30.2~mag on day $\approx $~4,190 (11.5~yrs) by \cite{Tucker2024}, using a distance modulus to M101 of $\mu=29.03$~mag \citep[6.4 Mpc;][]{Shappee2011}. 

The $3\sigma$ non-detection limit at 11.5~yrs rules out a surviving $\gtrsim 0.4\,\msun$ He WD donor, further constraining possible donor scenarios. We note that the shocked CO WD donor in \cite{Bhat2025} show a similarly fast brightening, which can be used in combination with our work to rule out a large parameter space for surviving degenerate companions for SN2011fe. While lower-mass He WD donors are still consistent with the non-detection limit, the amount of He mass loss increases, which may lead to stronger nebular phase emission lines. 
Tight limits on nebular emission line fluxes from stripped companion material have been placed for SN2011fe, though the determination of He mass loss remains uncertain and requires radiative transfer modeling \citep[][]{Lundqvist2015,Botyanszki2018}. Future studies will combine the non-detection limits of a surviving companion and companion mass loss, to determine the companion scenario for nearby SNe Ia like SN2011fe. 

Likewise, \cite{Do2021} placed a non-detection limit of 28~mag for SN1972E \citep[3.15 Mpc, $\mu$=27.49~mag;][]{Freedman2001} 33~yrs after explosion. This rules out any surviving He WD donor with $\iniMstar \gtrsim 0.3\,\msun$. 

Recently, \cite{Shields2022,Shields2023} conducted wide and deep searches for surviving, fast-moving companions (up to 2,500~$\kms$) in SN Ia remnants. For the Large Magellanic Cloud remnant SNR 0509-67.5 \citep[age~$\approx 300$~yrs;][]{Arunachalam2022}, \cite{Shields2023} found no star with unusual properties near the center. Their study was sensitive to stars $\gtrsim 0.2\,\rsun$ for $\Teff \approx 10^{4}$~K, and thus rule out all our He WD models at a similar age. Similarly, for the remnant of SN1006, \cite{Shields2022} found no high-proper motion star with $r$-band apparent magnitude brighter than $m_{r} = 21$, which is inconsistent with our models, which have $m_{r}$ brighter than 18 mag at this distance and age, unless the companion was ejected along our line-of-sight. In both cases, a surviving He WD companion can be ruled out unless there is strong localized dust \citep[][]{Shields2022}. 


\subsection{Comparison to observed hypervelocity stars}
\label{sec:comparison to observed hvs}

For comparison, on the HR diagram we show four hypervelocity stars whose inferred ejection velocities \citep[][]{ElBadry2023,Hollands2025} are consistent with a Roche lobe-filling He WD donor ($\lesssim $1,800\,$\kms$). 

D6-2 has an inferred ejection velocity 1,051$^{+62}_{-54}$\,$\kms$, consistent with a Roche lobe-filling $\approx 0.1-0.2\,\msun$ He WD \citep[][]{Bauer2021_D6,ElBadry2023}. It has well-measured $\Teff=$7,500~$\pm~100$~K and $R = 0.20$~$\pm~0.01~\rsun$ \citep[][]{Chandra2022}, an inferred age $\approx 10^{5}$~yrs based on its association with a known SN remnant \citep[][]{Shen2018_D6}, and is consistent with having a He-dominated atmosphere \citep[][]{Chandra2022}. 
\cite{Wong2024} showed an initially $0.126\,\msun$ high-entropy He WD (our \verb|0.126_3.9|) can reasonably reproduce its properties, albeit at a slightly older age $\approx 10^{6}$~yrs. Here we find that donors with slightly lower $\iniMstar \approx 0.08 - 0.10\,\msun$ (models \verb|0.08_2.7| \& \verb|0.10_3.9|), fit D6-2 even better. These models yield an age $\approx$~few $10^{5}$~yrs for D6-2, in excellent agreement with its inferred age. In addition, the final velocities of these models are $\approx 970 - $1,100~$\kms$, consistent with the inferred ejection velocity of D6-2. From these models, we infer a post-explosion mass $\approx 0.05\,\msun$ for D6-2, in rough agreement with $\approx 0.02\,\msun$ as \cite{Shen2025} inferred.

We estimate surface rotation periods $\approx 20 -24$~hrs at $10^{5}$~yrs for these two models, close to the $15.4$~hr periodic signal observed for D6-2 \citep[][]{Chandra2022}. This may suggest some internal angular momentum transport for D6-2. Donors that are slightly more compact yield a surface rotation period $\approx 14-17$~hrs, in better agreement with observations of D6-2. We note that there may exist hypervelocity stars redder than D6-2 \citep[see also][]{Shen2025}, for which we predict a significantly longer surface rotation period (e.g., $\approx60$~hrs for model \verb|0.08_3.9|), since these likely originate from lower-mass Roche lobe-filling donors that have longer orbital periods.

J1332-3541 has an inferred ejection velocity 1,619$^{+707}_{-320}\,\kms$, implying a Roche lobe-filling donor mass $\gtrsim 0.27\,\msun$ \citep[][]{ElBadry2023}. We adopt $\Teff =$~65,657~$\pm$~2,390~K \citep[][]{Werner2024} and $R = 0.017^{+0.013}_{-0.007}\,\rsun$ \citep[][]{ElBadry2023}. Although \cite{Werner2024} found a H-rich atmosphere, this may result from accretion of a tiny amount of interstellar medium \citep[][]{Shen2025}. Our models with $\iniMstar \approx 0.35 - 0.45\,\msun$ are consistent with the HR diagram location J1332-3541 at ages $\approx$~few~$10^{4} - 10^{5}$~yrs, as well as its ejection velocity. Our inferred age reasonably agrees with its inferred flight time from the midplane $\approx$~few~$10^{5}$~yrs \citep[][]{ElBadry2023}. 

The inferred ejection velocity of 1,864$^{+682}_{-416}\,\kms$ for J0546+0836 implies a donor mass $\gtrsim 0.34\,\msun$ \citep[][]{ElBadry2023}. We adopt $\Teff = $~95,000~$\pm$~15,000~K \citep[][]{Werner2024} and $R = 0.051^{+0.029}_{-0.021}\,\rsun$ \citep[][]{ElBadry2023}. 
We find that donors with $\iniMstar \approx 0.30 - 0.45\,\msun$ can explain its properties well at $\approx 10^{3} - 10^{4}$~yrs, with final velocities also consistent with its inferred ejection velocity. \cite{ElBadry2023} inferred a midplane flight time $\approx$~few~$10^{5}$~yrs, in slight tension with our inferred age. \cite{Werner2024} found a C/O-dominated atmosphere with a He mass fraction $\lesssim 0.05$, which hints at a C/O WD. However, if it is a He WD, its surface composition may be modified by burning during SN shock-heating, which may explain its current C/O-rich atmosphere \citep[see also][]{Papish2015}. This is very likely for the massive donors, $\iniMstar \approx 0.4\,\msun$, where we find that the upstream hemisphere of the donor can be shocked to $\rho \approx 10^{5}\,\gcc$ and $T\approx10^{9}$~K. The modified composition may change the opacity, and hence the thermal evolution of the shocked donor. However, cite{Bhat2025} do not find significant differences when including this effect in their CO WD models. In addition, nuclear energy release during the interaction may modify the thermal structure of the surviving donor leading to sustained burning \citep[e.g.,][]{Yamaguchi2025}, or even cause the donor to detonate \citep[e.g.,][]{Papish2015,Tanikawa2019,Boos2024}, which will be explored by future studies. 

We also include the recent discovered hypervelocity star candidate SDSS J1637+3631, which has an inferred ejection velocity $1870^{+360}_{-300}\,\kms$, $\Teff = 15680 \pm 250\,\rm{K}$, and $R = 0.114^{+0.019}_{-0.016} \, \rsun$ \citep[][]{Hollands2025}. Assuming a He WD origin, we infer a lower $\iniMstar \approx 0.15\,\msun$, which is similar to a final mass $\approx 0.18 - 0.20\,\msun$ inferred using \cite{Shen2025}. Both these are in tension with the spectroscopic fit $0.42^{+0.17}_{-0.15}\,\msun$ by \cite{Hollands2025}. 
Our $\iniMstar=0.15\,\msun$ model has a final velocity $\approx $1,$100\,\kms$, which is too low for the inferred ejection velocity of SDSS J1637+3631. 
From Figure \ref{fig:HR}, we infer an age $\approx$~few~$10^{5}$~yrs, which is an order of magnitude lower than the inferred flight time $4.5^{+0.4}_{-0.5}$~Myrs. 
In addition, \cite{Hollands2025} find a C/O-dominated atmosphere. Future work will explore whether this can be explained by heavy elements produced by surface nuclear burning during shock interaction.   
Overall, our models can explain several, but not all, of the observed properties of SDSS J1637+3631.

\bigskip

Our models can reasonably explain several hypervelocity stars, namely D6-2, J1332-3541 \& J0546+0836. We find that shocked donors with smaller $\iniMstar$ tend to have larger mass loss. In agreement with \cite{Zhang2019} \& \cite{Shen2025}, a hypervelocity star with a lower \textit{final} mass remains puffier for a longer amount of time. While \cite{Bhat2025} find that the shock-heating cannot account for the hypervelocity stars, this is likely a consequence of their choice of $\approx 0.5 - 1.0\,\msun$ WDs, which tend to cool quickly. Our work and \cite{Shen2025} suggest that a lower final mass $\lesssim 0.5\,\msun$ is required to explain the inflated state of the hypervelocity stars.

Our models do not account for the metal pollution from the ejecta bound to the donor, and simply assume solar metallicity $Z=0.014$ \citep[][]{Asplund2009}. 
As shown in Table \ref{tab:models}, the bound SN ejecta mass $M^{f}_{\rm ej}$ is negligible for our lowest-mass donors, but increases to $\approx 10^{-3}\,\msun$ for our most massive donors as the donor remains compact and has a stronger gravitational field. 
Additional metal pollution may result from surface nuclear burning during the shock interaction. 
Future studies will clarify the effects of gravitational settling and radiative levitation \citep[e.g.,][]{Zhang2019}.

\section{Conclusion}
\label{sec:conclusion}

We have studied the hydrodynamical interaction between SN Ia ejecta and Roche lobe-filling He WD companions with a wide range of initial masses $\iniMstar = 0.08 - 0.45\,\msun$, focusing on the donor mass loss and the SN impact on the donor. The surviving donor is flung off at roughly its orbital velocity, leading to hypervelocity stars. We improve upon the {\athena} simulations by \cite{Bauer2019} and \cite{Wong2023}, by employing a realistic EOS from {\mesa} that accounts for electron degeneracy, radiation, and Coulomb interactions, and accurately captures the temperature jump in the donor due to shock heating. 

We find a tight relation between the fractional mass loss of the companion, $\Delta \mstar / \mstar$ and the ratio between ejecta ram pressure and donor volume-average pressure, $\Pej / \iniPavg$ (see Section \ref{sec:mass loss}). When applied to Roche lobe-filling companions, this implies that a companion with lower mass and larger radius is more strongly impacted by SN ejecta \citep[see Section \ref{sec:theory} and ][]{Bauer2019,Wong2024}. A lower-mass Roche lobe-filling WD donor thus undergoes much stronger mass loss. Our lowest mass and most inflated donor, with initial mass $0.08\,\msun$, loses $\approx 0.06\,\msun$ ($\approx 70\%$ of its initial mass) and becomes significantly inflated due to strong shock-heating, whereas our most massive model with initial mass $0.45\,\msun$ loses only $\approx 0.005\,\msun$.

The tight relation between $\Delta \mstar / \mstar $ and $\Pej / \iniPavg$ is accurate to within a factor of $\approx 3$ across various companion stellar structures and ejecta structures (see Figure \ref{fig:delta_M} where we compare with \citealt{Marietta2000} \& \citealt{Hirai2018}). It can therefore be extended to any SN interaction with companion stars, like core-collapse SNe \citep[see also][]{Liu2015,Hirai2018,Cary2025} and SNe Iax \citep[see also][]{Liu2013a,Zeng2020}, motivating our providing a fitting function (Equation \ref{eqn:mass_loss}). 

We predict that the pure hydrodynamical impact of SNe Ia on \textit{Roche lobe-filling} CO WD companions leads to little mass loss (fractional mass loss $\lesssim 10^{-2}$), whereas \cite{Shen2025} found that explaining the HR diagram locations of the hypervelocity stars requires a much lower mass $\lesssim 0.5\,\msun$, than implied by their ejection velocities \citep[some require $\approx 1\,\msun$ assuming a Roche lobe-filling donor;][]{ElBadry2023}. This favors the interpretation that significant mass loss must have occurred \textit{before} the explosion from a partial merger \citep[][]{Shen2025}, during which the donor acquires a higher velocity from the plunge-in \citep[][]{Burmester2023,Glanz2024}. As the donor is now closer (higher ejecta ram pressure) and has lower central pressure, it will lose even more mass from SN impact \citep[e.g., Section \ref{sec:mass loss} \& ][]{Shen2025}.

Our predicted $\approx 0.005 - 0.06\,\msun$ of stripped He material, embedded within the ejecta, may lead to narrow emission lines in the nebular phase \citep[e.g.,][]{Mattila2005}. The typical velocity of the mass loss is roughly given by the escape velocity of the donor, increasing with donor mass, whereas the total amount of mass loss decreases with donor mass. The resulting mass loss ranges from $\approx$~1,000~$\kms$ ($\approx 0.06\,\msun$ mass loss) for our lowest mass donor, to $\approx$~4,600~$\kms$ ($\approx 0.005\,\msun$ mass loss) for our highest mass donor. Furthermore, mass loss with faster velocities are more mixed with SN ejecta. In order to meaningfully constrain any He-rich donor mass loss, we encourage future radiative transfer studies modeling nebular phase spectra to account for realistic velocity distributions and mixing.

While the donor receives some kick velocity from the ejecta up to $\approx 300\,\kms$, we emphasize that the final ejection velocity of the donor is almost entirely dictated by its orbital velocity $\gtrsim $~1,000~$\kms$, since the orbital and kick velocities add in quadrature and the former is much greater. 

Following the hydrodynamical interaction, we map the shocked donor into {\mesa} and model its subsequent stellar evolution. The shocked donor remains bright and later returns to the WD cooling track, on a timescale given by the heat diffusion timescale at the location of peak heating \citep[][]{Zhang2019,Bauer2019,Wong2024,Bhat2025}. We find reasonable agreement with four hypervelocity stars whose velocities are consistent with a He WD donor. Models with initial masses $\approx 0.08 - 0.10\,\msun$ and final masses $\approx 0.05\,\msun$ agree with the properties of D6-2 at an age $\approx \rm{few}~10^{5}$~yrs, consistent with its inferred age $\approx 10^{5}$~yrs from its association with a known SN remnant \citep[][]{Shen2018_D6}. We find that J0546+0836 and J1332-3541 are both consistent with a $\approx 0.3 - 0.45\,\msun$ He WD, while the HR diagram location of SDSS J1637+3631 can be explained by a $\approx 0.15\,\msun$ He WD. Both J0546+0836 and SDSS J1637+0836 have C/O-rich surfaces, but their surfaces may have been modified by nuclear burning during shock interaction and accretion of SN ejecta. While \cite{Bhat2025} find that shock heating cannot explain the hypervelocity WDs, their $0.5-1.0\,\msun$ WDs are much more massive than ours. These more massive WDs release the deposited shock energy, and hence return to the WD cooling track, more quickly. In contrast, our He WDs which have lower masses are able to remain inflated for much longer, consistent with theoretical predictions \citep[][]{Zhang2019,Shen2025}. 

Due to heat diffusion, shocked donors with initial masses $\gtrsim 0.15\,\msun$ brighten and expand significantly on a timescale ranging from $\approx 1$~yr for a $0.45\,\msun$ donor, to $\approx 10^{4}$~yrs for $0.15\,\msun$ donor. 
We find that surviving He WD donors $\gtrsim 0.4\,\msun$ and $\gtrsim 0.3 \,\msun$ can be ruled out for SN2011fe and SN1972E respectively, by comparing to deep non-detection limits by \cite{Tucker2024} \& \cite{Do2021}. Since higher-mass shocked donors brighten even faster \citep[][]{Zhang2019}, a surviving massive WD donor such as those in \cite{Bhat2025} \& \cite{Glanz2024} can also be ruled out for these SNe. While very strict limits have already been placed on nondegenerate donors \citep[][]{Tucker2024}, our study now allows constraints on \textit{surviving} He WD companions. Importantly, obtaining deep imaging of historical SNe Ia in nearby ($\lesssim 10$~Mpc) galaxies can be a powerful tool to test the existence of any surviving companions \citep[][]{Do2021,Tucker2024}. 

Moreover, we find that all our models are inconsistent with the limits placed on surviving companions for SNR 0509-67.5 \citep[][]{Shields2023} and SN1006 \citep[][]{Shields2022}, unless there is strong localized extinction. This either rules out a He WD companion scenario, or that there is no surviving companion. 

Recent studies have suggested that the donor may undergo a detonation when shocked by the ejecta, which nicely explains the lack of surviving companions in SN Ia remnants \citep[][]{Tanikawa2019,Pakmor2022,Shen2024,Boos2024}. 
Our early assessment of He fusion during shock traversal indicates that stars less massive than $\approx 0.3\,\msun$ will not experience any He burning adequate to change the entropy of the shock. Future studies will account for nuclear burning and test whether the donor undergoes a detonation. Alternatively, a surviving donor that has undergone strong nuclear burning during the interaction may be masked by heavy elements and its thermal profile may be different than modeled here. We leave these interesting questions to future explorations.

\section*{Acknowledgments}

We thank the referee for their constructive suggestions that have greatly improved our manuscript. 
We are grateful to Aakash Bhat, Kareem El-Badry, Jim Fuller, Logan Prust and Ken Shen for stimulating conversations about the hypervelocity stars. 
We thank Andy Howell for productive conversations about SNe Ia observations.
We thank Joshua Shields for helpful discussion on searches for surviving companions in SN remnants. 
We appreciate Chris White's help and guidance in setting up the {\athena} simulations, Matthew Coleman's effort in developing the general EOS capability in {\athena}, and other {\athena} developers who provided guidance on debugging and running simulations. 
This work was supported, in part, by the National Science Foundation through grant PHY-2309135, and by the Gordon and Betty Moore Foundation through grant GBMF5076. 
This work used the Expanse cluster at the San Diego Supercomputer Center (\href{doi.org/10.1145/3437359.3465588}{doi.org/10.1145/3437359.3465588}) through allocation PHY240315 from the Advanced Cyberinfrastructure Coordination Ecosystem: Services \& Support (ACCESS) program (\href{doi.org/10.1145/3569951.3597559}{doi.org/10.1145/3569951.3597559}), which is supported by U.S. National Science Foundation grants \#2138259, \#2138286, \#2138307, \#2137603, and \#2138296.
Earlier preliminary runs were performed at facilities supported by the Scientific Computing Core at the Flatiron Institute, a division of the Simons Foundation, and at computational facilities purchased with funds from the National Science Foundation (CNS-1725797) and administered by the Center for Scientific Computing (CSC). The CSC is supported by the California NanoSystems Institute and the Materials Research Science and Engineering Center (MRSEC; NSF DMR 2308708) at UC Santa Barbara.

\software{
\texttt{Athena++} \citep[git-version 97d58f1;][]{Stone2020}, 
\texttt{MESA} \citep[v24.03.1;][]{MESAI,MESAII,MESAIII,MESAIV,MESAV,MESAVI}, 
\texttt{py\_mesa\_reader} \citep{bill_wolf_2017_826958},
\texttt{ipython/jupyter} \citep{perez_2007_aa,kluyver_2016_aa},
\texttt{matplotlib} \citep{hunter_2007_aa},
\texttt{NumPy} \citep{numpy2020}, 
\texttt{SciPy} \citep{scipy2020}, 
\texttt{Astropy} \citep{astropy2013,astropy2018,astropy2022},
and 
\texttt{Python} from \href{https://www.python.org}{python.org}
}

\newpage

\appendix

\twocolumngrid

\section{EOS table}
\label{sec:EOS_table}

We generate a table of $P(\rho,e)$, $e(\rho,P)$ and $c_{\rm s} (\rho, P)$, where $P$ is the total pressure, $\rho$ is the gas density, $e$ is the internal energy density, and $c_{\rm s}$ is the adiabatic sound speed. 
This table is based on the {\mesa} EOS, which is a blend of the OPAL \citep{Rogers2002}, SCVH
\citep{Saumon1995}, FreeEOS \citep{Irwin2004}, HELM \citep{Timmes2000}, 
PC \citep{Potekhin2010}, and Skye \citep{Jermyn2021} EOSes. 

Since the {\mesa} EOS uses $(\rho,T)$ as base variables, we iterate to solve for $T$ given, say, a $(\rho,P)$ pair, and subsequently obtain $e$. 
Under strongly degenerate conditions, $P$ and $e$ vary weakly with $T$, so we generate a table particularly dense in $P$ and $e$ to ensure accuracy. 
Our EOS table spans 2501 points from $\log_{10}\rho = -3$ to 7, and 160001 points from $\log_{10} P $ \& $\log_{10} e = 12.4$ to $21.0$, and is regular in logarithmic space in the base variables. 
We test the accuracy of our EOS table as follows. We apply the consecutive operations $e_{\rm interp} = e(P,\rho)$ and then $P_{\rm interp} = P(e_{\rm interp}, \rho)$, and compare between $P_{\rm interp}$ and the true $P$. In general we find errors less than $10^{-6}$. For the central conditions of our $0.45\,\msun$ He WD model, $\rho \approx 10^{6}\,\gcc$ and $T \approx 10^{7}$~K, this corresponds to $T$ errors less than 5\%. 

We tested the accuracy of the tabulated EOS method with the $0.45\,\msun$ He WD, our most degenerate model. We performed two simulations, (i) one with a tabulated EOS constructed from the HELM EOS with the aforementioned dimensions, and (ii) one coupled to the HELM EOS on-the-fly, where at each timestep, $T$ is solved iteratively until we achieve a fractional error $\lesssim 10^{-8}$ in $P$ or $e$. While method (ii) requires a runtime 20 times longer on 60 cores, we use it to benchmark the accuracy of the tabulated EOS method. 
SN ejecta is injected immediately after the $0.45\,\msun$ He WD is mapped onto the grid, and we continue until shock breakout from the downstream side of the donor. 
The evolution of the two runs are nearly identical. 
Before the shock reaches the center, the difference in $\log_{10} T_{\rm c}$ is $\approx 10^{-2}$ ($\approx 2 \%$ error in $T_{\rm c}$), and drops to $\approx 10^{-3}$ ($\approx 0.2\%$  error in $T_{\rm c}$) post-shock, while the difference in $\log_{10} \rhoc$ is $\lesssim 10^{-5}$. We conclude that our EOS table is sufficiently accurate, given that a more accurate table requires a denser table and hence a steep trade-off in memory storage (our EOS table stored in hdf5 format requires 12 GB). 

Finally, for this work we assume a pure He composition. It only loses accuracy in the unshocked, gas-pressure dominated ejecta, and is therefore sufficient for our purposes. Future work will allow for a varying composition. 


\bibliography{athena,main,mesa,software}{}

\begin{thebibliography}{}
\expandafter\ifx\csname natexlab\endcsname\relax\def\natexlab#1{#1}\fi
\providecommand{\url}[1]{\href{#1}{#1}}
\providecommand{\dodoi}[1]{doi:~\href{http://doi.org/#1}{\nolinkurl{#1}}}
\providecommand{\doeprint}[1]{\href{http://ascl.net/#1}{\nolinkurl{http://ascl.net/#1}}}
\providecommand{\doarXiv}[1]{\href{https://arxiv.org/abs/#1}{\nolinkurl{https://arxiv.org/abs/#1}}}

\bibitem[{P. {Arunachalam} {et~al.}(2022){Arunachalam}, {Hughes}, {Hovey}, \& {Eriksen}}]{Arunachalam2022}
{Arunachalam}, P., {Hughes}, J.~P., {Hovey}, L., \& {Eriksen}, K. 2022, \bibinfo{title}{{A Hydro-based MCMC Analysis of SNR 0509-67.5: Revealing the Explosion Properties from Fluid Discontinuities Alone},} \apj, 938, 121, \dodoi{10.3847/1538-4357/ac927c}

\bibitem[{M. {Asplund} {et~al.}(2009){Asplund}, {Grevesse}, {Sauval}, \& {Scott}}]{Asplund2009}
{Asplund}, M., {Grevesse}, N., {Sauval}, A.~J., \& {Scott}, P. 2009, \bibinfo{title}{{The Chemical Composition of the Sun},} \araa, 47, 481, \dodoi{10.1146/annurev.astro.46.060407.145222}

\bibitem[{ {Astropy Collaboration} {et~al.}(2013){Astropy Collaboration}, {Robitaille}, {Tollerud}, {Greenfield}, {Droettboom}, {Bray}, {Aldcroft}, {Davis}, {Ginsburg}, {Price-Whelan}, {Kerzendorf}, {Conley}, {Crighton}, {Barbary}, {Muna}, {Ferguson}, {Grollier}, {Parikh}, {Nair}, {Unther}, {Deil}, {Woillez}, {Conseil}, {Kramer}, {Turner}, {Singer}, {Fox}, {Weaver}, {Zabalza}, {Edwards}, {Azalee Bostroem}, {Burke}, {Casey}, {Crawford}, {Dencheva}, {Ely}, {Jenness}, {Labrie}, {Lim}, {Pierfederici}, {Pontzen}, {Ptak}, {Refsdal}, {Servillat}, \& {Streicher}}]{astropy2013}
{Astropy Collaboration}, {Robitaille}, T.~P., {Tollerud}, E.~J., {et~al.} 2013, \bibinfo{title}{{Astropy: A community Python package for astronomy},} \aap, 558, A33, \dodoi{10.1051/0004-6361/201322068}

\bibitem[{ {Astropy Collaboration} {et~al.}(2018){Astropy Collaboration}, {Price-Whelan}, {Sip{\H{o}}cz}, {G{\"u}nther}, {Lim}, {Crawford}, {Conseil}, {Shupe}, {Craig}, {Dencheva}, {Ginsburg}, {VanderPlas}, {Bradley}, {P{\'e}rez-Su{\'a}rez}, {de Val-Borro}, {Aldcroft}, {Cruz}, {Robitaille}, {Tollerud}, {Ardelean}, {Babej}, {Bach}, {Bachetti}, {Bakanov}, {Bamford}, {Barentsen}, {Barmby}, {Baumbach}, {Berry}, {Biscani}, {Boquien}, {Bostroem}, {Bouma}, {Brammer}, {Bray}, {Breytenbach}, {Buddelmeijer}, {Burke}, {Calderone}, {Cano Rodr{\'\i}guez}, {Cara}, {Cardoso}, {Cheedella}, {Copin}, {Corrales}, {Crichton}, {D'Avella}, {Deil}, {Depagne}, {Dietrich}, {Donath}, {Droettboom}, {Earl}, {Erben}, {Fabbro}, {Ferreira}, {Finethy}, {Fox}, {Garrison}, {Gibbons}, {Goldstein}, {Gommers}, {Greco}, {Greenfield}, {Groener}, {Grollier}, {Hagen}, {Hirst}, {Homeier}, {Horton}, {Hosseinzadeh}, {Hu}, {Hunkeler}, {Ivezi{\'c}}, {Jain}, {Jenness}, {Kanarek}, {Kendrew}, {Kern}, {Kerzendorf}, {Khvalko}, {King}, {Kirkby}, {Kulkarni},
  {Kumar}, {Lee}, {Lenz}, {Littlefair}, {Ma}, {Macleod}, {Mastropietro}, {McCully}, {Montagnac}, {Morris}, {Mueller}, {Mumford}, {Muna}, {Murphy}, {Nelson}, {Nguyen}, {Ninan}, {N{\"o}the}, {Ogaz}, {Oh}, {Parejko}, {Parley}, {Pascual}, {Patil}, {Patil}, {Plunkett}, {Prochaska}, {Rastogi}, {Reddy Janga}, {Sabater}, {Sakurikar}, {Seifert}, {Sherbert}, {Sherwood-Taylor}, {Shih}, {Sick}, {Silbiger}, {Singanamalla}, {Singer}, {Sladen}, {Sooley}, {Sornarajah}, {Streicher}, {Teuben}, {Thomas}, {Tremblay}, {Turner}, {Terr{\'o}n}, {van Kerkwijk}, {de la Vega}, {Watkins}, {Weaver}, {Whitmore}, {Woillez}, {Zabalza}, \& {Astropy Contributors}}]{astropy2018}
{Astropy Collaboration}, {Price-Whelan}, A.~M., {Sip{\H{o}}cz}, B.~M., {et~al.} 2018, \bibinfo{title}{{The Astropy Project: Building an Open-science Project and Status of the v2.0 Core Package},} \aj, 156, 123, \dodoi{10.3847/1538-3881/aabc4f}

\bibitem[{ {Astropy Collaboration} {et~al.}(2022){Astropy Collaboration}, {Price-Whelan}, {Lim}, {Earl}, {Starkman}, {Bradley}, {Shupe}, {Patil}, {Corrales}, {Brasseur}, {N{\"o}the}, {Donath}, {Tollerud}, {Morris}, {Ginsburg}, {Vaher}, {Weaver}, {Tocknell}, {Jamieson}, {van Kerkwijk}, {Robitaille}, {Merry}, {Bachetti}, {G{\"u}nther}, {Aldcroft}, {Alvarado-Montes}, {Archibald}, {B{\'o}di}, {Bapat}, {Barentsen}, {Baz{\'a}n}, {Biswas}, {Boquien}, {Burke}, {Cara}, {Cara}, {Conroy}, {Conseil}, {Craig}, {Cross}, {Cruz}, {D'Eugenio}, {Dencheva}, {Devillepoix}, {Dietrich}, {Eigenbrot}, {Erben}, {Ferreira}, {Foreman-Mackey}, {Fox}, {Freij}, {Garg}, {Geda}, {Glattly}, {Gondhalekar}, {Gordon}, {Grant}, {Greenfield}, {Groener}, {Guest}, {Gurovich}, {Handberg}, {Hart}, {Hatfield-Dodds}, {Homeier}, {Hosseinzadeh}, {Jenness}, {Jones}, {Joseph}, {Kalmbach}, {Karamehmetoglu}, {Ka{\l}uszy{\'n}ski}, {Kelley}, {Kern}, {Kerzendorf}, {Koch}, {Kulumani}, {Lee}, {Ly}, {Ma}, {MacBride}, {Maljaars}, {Muna}, {Murphy}, {Norman},
  {O'Steen}, {Oman}, {Pacifici}, {Pascual}, {Pascual-Granado}, {Patil}, {Perren}, {Pickering}, {Rastogi}, {Roulston}, {Ryan}, {Rykoff}, {Sabater}, {Sakurikar}, {Salgado}, {Sanghi}, {Saunders}, {Savchenko}, {Schwardt}, {Seifert-Eckert}, {Shih}, {Jain}, {Shukla}, {Sick}, {Simpson}, {Singanamalla}, {Singer}, {Singhal}, {Sinha}, {Sip{\H{o}}cz}, {Spitler}, {Stansby}, {Streicher}, {{\v{S}}umak}, {Swinbank}, {Taranu}, {Tewary}, {Tremblay}, {de Val-Borro}, {Van Kooten}, {Vasovi{\'c}}, {Verma}, {de Miranda Cardoso}, {Williams}, {Wilson}, {Winkel}, {Wood-Vasey}, {Xue}, {Yoachim}, {Zhang}, {Zonca}, \& {Astropy Project Contributors}}]{astropy2022}
{Astropy Collaboration}, {Price-Whelan}, A.~M., {Lim}, P.~L., {et~al.} 2022, \bibinfo{title}{{The Astropy Project: Sustaining and Growing a Community-oriented Open-source Project and the Latest Major Release (v5.0) of the Core Package},} \apj, 935, 167, \dodoi{10.3847/1538-4357/ac7c74}

\bibitem[{E.~B. {Bauer} {et~al.}(2021){Bauer}, {Chandra}, {Shen}, \& {Hermes}}]{Bauer2021_D6}
{Bauer}, E.~B., {Chandra}, V., {Shen}, K.~J., \& {Hermes}, J.~J. 2021, \bibinfo{title}{{Masses of White Dwarf Binary Companions to Type Ia Supernovae Measured from Runaway Velocities},} \apjl, 923, L34, \dodoi{10.3847/2041-8213/ac432d}

\bibitem[{E.~B. {Bauer} {et~al.}(2019){Bauer}, {White}, \& {Bildsten}}]{Bauer2019}
{Bauer}, E.~B., {White}, C.~J., \& {Bildsten}, L. 2019, \bibinfo{title}{{Remnants of Subdwarf Helium Donor Stars Ejected from Close Binaries with Thermonuclear Supernovae},} \apj, 887, 68, \dodoi{10.3847/1538-4357/ab4ea4}

\bibitem[{A. {Bhat} {et~al.}(2025){Bhat}, {Bauer}, {Pakmor}, {Shen}, {Caiazzo}, {Rajamuthukumar}, {El-Badry}, \& {Kerzendorf}}]{Bhat2025}
{Bhat}, A., {Bauer}, E.~B., {Pakmor}, R., {et~al.} 2025, \bibinfo{title}{{Supernova shocks cannot explain the inflated state of hypervelocity runaways from white dwarf binaries},} \aap, 693, A114, \dodoi{10.1051/0004-6361/202451371}

\bibitem[{L. {Bildsten} {et~al.}(2007){Bildsten}, {Shen}, {Weinberg}, \& {Nelemans}}]{Bildsten2007}
{Bildsten}, L., {Shen}, K.~J., {Weinberg}, N.~N., \& {Nelemans}, G. 2007, \bibinfo{title}{{Faint Thermonuclear Supernovae from AM Canum Venaticorum Binaries},} \apjl, 662, L95, \dodoi{10.1086/519489}

\bibitem[{P. {Boehner} {et~al.}(2017){Boehner}, {Plewa}, \& {Langer}}]{Boehner2017}
{Boehner}, P., {Plewa}, T., \& {Langer}, N. 2017, \bibinfo{title}{{Imprints of the ejecta-companion interaction in Type Ia supernovae: main-sequence, subgiant, and red giant companions},} \mnras, 465, 2060, \dodoi{10.1093/mnras/stw2737}

\bibitem[{S.~J. {Boos} {et~al.}(2024){Boos}, {Townsley}, \& {Shen}}]{Boos2024}
{Boos}, S.~J., {Townsley}, D.~M., \& {Shen}, K.~J. 2024, \bibinfo{title}{{Type Ia Supernovae Can Arise from the Detonations of Both Stars in a Double Degenerate Binary},} arXiv e-prints, arXiv:2401.08011, \dodoi{10.48550/arXiv.2401.08011}

\bibitem[{J. {Boty{\'a}nszki} {et~al.}(2018){Boty{\'a}nszki}, {Kasen}, \& {Plewa}}]{Botyanszki2018}
{Boty{\'a}nszki}, J., {Kasen}, D., \& {Plewa}, T. 2018, \bibinfo{title}{{Multidimensional Models of Type Ia Supernova Nebular Spectra: Strong Emission Lines from Stripped Companion Gas Rule Out Classic Single-degenerate Systems},} \apjl, 852, L6, \dodoi{10.3847/2041-8213/aaa07b}

\bibitem[{J. {Braudo} \& N. {Soker}(2024){Braudo} \& {Soker}}]{Braudo2024}
{Braudo}, J., \& {Soker}, N. 2024, \bibinfo{title}{{The runaway velocity of the white dwarf companion in the double detonation scenario of supernovae},} The Open Journal of Astrophysics, 7, 7, \dodoi{10.21105/astro.2310.16554}

\bibitem[{U.~P. {Burmester} {et~al.}(2023){Burmester}, {Ferrario}, {Pakmor}, {Seitenzahl}, {Ruiter}, \& {Hole}}]{Burmester2023}
{Burmester}, U.~P., {Ferrario}, L., {Pakmor}, R., {et~al.} 2023, \bibinfo{title}{{AREPO white dwarf merger simulations resulting in edge-lit detonation and run-away hypervelocity companion},} \mnras, 523, 527, \dodoi{10.1093/mnras/stad1394}

\bibitem[{S. {Cary} {et~al.}(2025){Cary}, {Lu}, {Leung}, \& {Wong}}]{Cary2025}
{Cary}, S., {Lu}, W., {Leung}, C., \& {Wong}, T. L.~S. 2025, \bibinfo{title}{{Accretion from a Shock-Inflated Companion: Spinning Down Neutron Stars to Hour-Long Periods},} arXiv e-prints, arXiv:2507.10682, \dodoi{10.48550/arXiv.2507.10682}

\bibitem[{V. {Chandra} {et~al.}(2022){Chandra}, {Hwang}, {Zakamska}, {Blouin}, {Swan}, {Marsh}, {Shen}, {G{\"a}nsicke}, {Hermes}, {Putterman}, {Bauer}, {Petrosky}, {Dhillon}, {Littlefair}, \& {Ashley}}]{Chandra2022}
{Chandra}, V., {Hwang}, H.-C., {Zakamska}, N.~L., {et~al.} 2022, \bibinfo{title}{{The SN Ia runaway LP 398-9: detection of circumstellar material and surface rotation},} \mnras, 512, 6122, \dodoi{10.1093/mnras/stac883}

\bibitem[{M.~S.~B. {Coleman}(2020){Coleman}}]{Coleman2020}
{Coleman}, M. S.~B. 2020, \bibinfo{title}{{An Extension of the Athena++ Framework for General Equations of State},} \apjs, 248, 7, \dodoi{10.3847/1538-4365/ab82ff}

\bibitem[{L. {Dessart} {et~al.}(2012){Dessart}, {Hillier}, {Li}, \& {Woosley}}]{Dessart2012}
{Dessart}, L., {Hillier}, D.~J., {Li}, C., \& {Woosley}, S. 2012, \bibinfo{title}{{On the nature of supernovae Ib and Ic},} \mnras, 424, 2139, \dodoi{10.1111/j.1365-2966.2012.21374.x}

\bibitem[{L. {Dessart} {et~al.}(2020){Dessart}, {Leonard}, \& {Prieto}}]{Dessart2020_H}
{Dessart}, L., {Leonard}, D.~C., \& {Prieto}, J.~L. 2020, \bibinfo{title}{{Spectral signatures of H-rich material stripped from a non-degenerate companion by a Type Ia supernova},} \aap, 638, A80, \dodoi{10.1051/0004-6361/202037854}

\bibitem[{G. {Dimitriadis} {et~al.}(2019){Dimitriadis}, {Rojas-Bravo}, {Kilpatrick}, {Foley}, {Piro}, {Brown}, {Guhathakurta}, {Quirk}, {Rest}, {Strampelli}, {Tucker}, \& {Villar}}]{Dimitriadis2019}
{Dimitriadis}, G., {Rojas-Bravo}, C., {Kilpatrick}, C.~D., {et~al.} 2019, \bibinfo{title}{{Nebular Spectroscopy of Kepler{\textquoteright}s Brightest Supernova},} \apjl, 870, L14, \dodoi{10.3847/2041-8213/aaf9b1}

\bibitem[{A. {Do} {et~al.}(2021){Do}, {Shappee}, {De Cuyper}, {Tonry}, {Hunt}, {Schweizer}, {Phillips}, {Burns}, {Beaton}, \& {Hainaut}}]{Do2021}
{Do}, A., {Shappee}, B.~J., {De Cuyper}, J.-P., {et~al.} 2021, \bibinfo{title}{{Blast from the past: constraining progenitor models of SN 1972E},} \mnras, 508, 3649, \dodoi{10.1093/mnras/stab2660}

\bibitem[{K. {El-Badry} {et~al.}(2023){El-Badry}, {Shen}, {Chandra}, {Bauer}, {Fuller}, {Strader}, {Chomiuk}, {Naidu}, {Caiazzo}, {Rodriguez}, {Nagarajan}, {Yamaguchi}, {Vanderbosch}, {Roulston}, {G{\"a}nsicke}, {Han}, {Burdge}, {Filippenko}, {Brink}, \& {Zheng}}]{ElBadry2023}
{El-Badry}, K., {Shen}, K.~J., {Chandra}, V., {et~al.} 2023, \bibinfo{title}{{The fastest stars in the Galaxy},} The Open Journal of Astrophysics, 6, 28, \dodoi{10.21105/astro.2306.03914}

\bibitem[{C. {Feldman} {et~al.}(2024){Feldman}, {Eisenberg}, {Townsley}, \& {Calder}}]{Feldman2024}
{Feldman}, C., {Eisenberg}, E., {Townsley}, D.~M., \& {Calder}, A.~C. 2024, in Journal of Physics Conference Series, Vol. 2742, Journal of Physics Conference Series (IOP), 012022, \dodoi{10.1088/1742-6596/2742/1/012022}

\bibitem[{W.~L. {Freedman} {et~al.}(2001){Freedman}, {Madore}, {Gibson}, {Ferrarese}, {Kelson}, {Sakai}, {Mould}, {Kennicutt}, {Ford}, {Graham}, {Huchra}, {Hughes}, {Illingworth}, {Macri}, \& {Stetson}}]{Freedman2001}
{Freedman}, W.~L., {Madore}, B.~F., {Gibson}, B.~K., {et~al.} 2001, \bibinfo{title}{{Final Results from the Hubble Space Telescope Key Project to Measure the Hubble Constant},} \apj, 553, 47, \dodoi{10.1086/320638}

\bibitem[{H. {Glanz} {et~al.}(2024){Glanz}, {Perets}, {Bhat}, \& {Pakmor}}]{Glanz2024}
{Glanz}, H., {Perets}, H.~B., {Bhat}, A., \& {Pakmor}, R. 2024, \bibinfo{title}{{Origins of the fastest stars from merger-disruption of He-CO white dwarfs},} arXiv e-prints, arXiv:2410.17306, \dodoi{10.48550/arXiv.2410.17306}

\bibitem[{M.~L. {Graham} {et~al.}(2017){Graham}, {Kumar}, {Hosseinzadeh}, {Hiramatsu}, {Arcavi}, {Howell}, {Valenti}, {Sand}, {Parrent}, {McCully}, \& {Filippenko}}]{Graham2017}
{Graham}, M.~L., {Kumar}, S., {Hosseinzadeh}, G., {et~al.} 2017, \bibinfo{title}{{Nebular-phase spectra of nearby Type Ia Supernovae},} \mnras, 472, 3437, \dodoi{10.1093/mnras/stx2224}

\bibitem[{M.~J. {Green} {et~al.}(2025){Green}, {van Roestel}, \& {Wong}}]{Green2025}
{Green}, M.~J., {van Roestel}, J., \& {Wong}, T. L.~S. 2025, \bibinfo{title}{{A Catalogue of Ultracompact Mass-Transferring White Dwarf Binaries},} arXiv e-prints, arXiv:2505.10535, \dodoi{10.48550/arXiv.2505.10535}

\bibitem[{C.~R. {Harris} {et~al.}(2020){Harris}, {Millman}, {van der Walt}, {Gommers}, {Virtanen}, {Cournapeau}, {Wieser}, {Taylor}, {Berg}, {Smith}, {Kern}, {Picus}, {Hoyer}, {van Kerkwijk}, {Brett}, {Haldane}, {del R{\'\i}o}, {Wiebe}, {Peterson}, {G{\'e}rard-Marchant}, {Sheppard}, {Reddy}, {Weckesser}, {Abbasi}, {Gohlke}, \& {Oliphant}}]{numpy2020}
{Harris}, C.~R., {Millman}, K.~J., {van der Walt}, S.~J., {et~al.} 2020, \bibinfo{title}{{Array programming with NumPy},} \nat, 585, 357, \dodoi{10.1038/s41586-020-2649-2}

\bibitem[{R. {Hirai} {et~al.}(2018){Hirai}, {Podsiadlowski}, \& {Yamada}}]{Hirai2018}
{Hirai}, R., {Podsiadlowski}, P., \& {Yamada}, S. 2018, \bibinfo{title}{{Comprehensive Study of Ejecta-companion Interaction for Core-collapse Supernovae in Massive Binaries},} \apj, 864, 119, \dodoi{10.3847/1538-4357/aad6a0}

\bibitem[{M.~A. {Hollands} {et~al.}(2025){Hollands}, {Shen}, {Raddi}, {G{\"a}nsicke}, {Bauer}, \& {Rebassa-Mansergas}}]{Hollands2025}
{Hollands}, M.~A., {Shen}, K.~J., {Raddi}, R., {et~al.} 2025, \bibinfo{title}{{Spectroscopic and kinematic analyses of a warm survivor of a D$^{6}$ supernova},} \mnras, \dodoi{10.1093/mnras/staf950}

\bibitem[{G. {Hosseinzadeh} {et~al.}(2022){Hosseinzadeh}, {Sand}, {Lundqvist}, {Andrews}, {Bostroem}, {Dong}, {Janzen}, {Jencson}, {Lundquist}, {Meza Retamal}, {Pearson}, {Valenti}, {Wyatt}, {Burke}, {Howell}, {McCully}, {Newsome}, {Gonzalez}, {Pellegrino}, {Terreran}, {Kwok}, {Jha}, {Strader}, {Kundu}, {Ryder}, {Haislip}, {Kouprianov}, \& {Reichart}}]{Hosseinzadeh2022}
{Hosseinzadeh}, G., {Sand}, D.~J., {Lundqvist}, P., {et~al.} 2022, \bibinfo{title}{{Constraining the Progenitor System of the Type Ia Supernova 2021aefx},} \apjl, 933, L45, \dodoi{10.3847/2041-8213/ac7cef}

\bibitem[{J.~D. Hunter(2007)Hunter}]{hunter_2007_aa}
Hunter, J.~D. 2007, \bibinfo{title}{Matplotlib: A 2D graphics environment,} Computing In Science \&amp; Engineering, 9, 90

\bibitem[{A.~W. {Irwin}(2004){Irwin}}]{Irwin2004}
{Irwin}, A.~W. 2004, \bibinfo{title}{The FreeEOS Code for Calculating the Equation of State for Stellar Interiors,} \url{http://freeeos.sourceforge.net/}

\bibitem[{W.~V. {Jacobson-Gal{\'a}n} {et~al.}(2019){Jacobson-Gal{\'a}n}, {Foley}, {Schwab}, {Dimitriadis}, {Dong}, {Jha}, {Kasen}, {Kilpatrick}, \& {Thomas}}]{JacobsonGalan2019}
{Jacobson-Gal{\'a}n}, W.~V., {Foley}, R.~J., {Schwab}, J., {et~al.} 2019, \bibinfo{title}{{Detection of circumstellar helium in Type Iax progenitor systems},} \mnras, 487, 2538, \dodoi{10.1093/mnras/stz1305}

\bibitem[{A.~S. {Jermyn} {et~al.}(2021){Jermyn}, {Schwab}, {Bauer}, {Timmes}, \& {Potekhin}}]{Jermyn2021}
{Jermyn}, A.~S., {Schwab}, J., {Bauer}, E., {Timmes}, F.~X., \& {Potekhin}, A.~Y. 2021, \bibinfo{title}{{Skye: A Differentiable Equation of State},} \apj, 913, 72, \dodoi{10.3847/1538-4357/abf48e}

\bibitem[{A.~S. {Jermyn} {et~al.}(2023){Jermyn}, {Bauer}, {Schwab}, {Farmer}, {Ball}, {Bellinger}, {Dotter}, {Joyce}, {Marchant}, {Mombarg}, {Wolf}, {Sunny Wong}, {Cinquegrana}, {Farrell}, {Smolec}, {Thoul}, {Cantiello}, {Herwig}, {Toloza}, {Bildsten}, {Townsend}, \& {Timmes}}]{MESAVI}
{Jermyn}, A.~S., {Bauer}, E.~B., {Schwab}, J., {et~al.} 2023, \bibinfo{title}{{Modules for Experiments in Stellar Astrophysics (MESA): Time-dependent Convection, Energy Conservation, Automatic Differentiation, and Infrastructure},} \apjs, 265, 15, \dodoi{10.3847/1538-4365/acae8d}

\bibitem[{D. {Kasen}(2010){Kasen}}]{Kasen2010}
{Kasen}, D. 2010, \bibinfo{title}{{Seeing the Collision of a Supernova with Its Companion Star},} \apj, 708, 1025, \dodoi{10.1088/0004-637X/708/2/1025}

\bibitem[{T. Kluyver {et~al.}(2016)Kluyver, Ragan-Kelley, P{\'e}rez, Granger, Bussonnier, Frederic, Kelley, Hamrick, Grout, Corlay, {et~al.}}]{kluyver_2016_aa}
Kluyver, T., Ragan-Kelley, B., P{\'e}rez, F., {et~al.} 2016, in Positioning and Power in Academic Publishing: Players, Agents and Agendas: Proceedings of the 20th International Conference on Electronic Publishing, IOS Press, 87

\bibitem[{I. {Linial} \& B.~D. {Metzger}(2023){Linial} \& {Metzger}}]{Linial2023}
{Linial}, I., \& {Metzger}, B.~D. 2023, \bibinfo{title}{{EMRI + TDE = QPE: Periodic X-Ray Flares from Star-Disk Collisions in Galactic Nuclei},} \apj, 957, 34, \dodoi{10.3847/1538-4357/acf65b}

\bibitem[{Z.-W. {Liu} {et~al.}(2013{\natexlab{a}}){Liu}, {Kromer}, {Fink}, {Pakmor}, {R{\"o}pke}, {Chen}, {Wang}, \& {Han}}]{Liu2013a}
{Liu}, Z.-W., {Kromer}, M., {Fink}, M., {et~al.} 2013{\natexlab{a}}, \bibinfo{title}{{Predicting the Amount of Hydrogen Stripped by the SN Explosion for SN 2002cx-like SNe Ia},} \apj, 778, 121, \dodoi{10.1088/0004-637X/778/2/121}

\bibitem[{Z.~W. {Liu} {et~al.}(2012){Liu}, {Pakmor}, {R{\"o}pke}, {Edelmann}, {Wang}, {Kromer}, {Hillebrandt}, \& {Han}}]{Liu2012}
{Liu}, Z.~W., {Pakmor}, R., {R{\"o}pke}, F.~K., {et~al.} 2012, \bibinfo{title}{{Three-dimensional simulations of the interaction between Type Ia supernova ejecta and their main sequence companions},} \aap, 548, A2, \dodoi{10.1051/0004-6361/201219357}

\bibitem[{Z.-W. {Liu} {et~al.}(2023){Liu}, {R{\"o}pke}, \& {Han}}]{Liu2023_review}
{Liu}, Z.-W., {R{\"o}pke}, F.~K., \& {Han}, Z. 2023, \bibinfo{title}{{Type Ia Supernova Explosions in Binary Systems: A Review},} Research in Astronomy and Astrophysics, 23, 082001, \dodoi{10.1088/1674-4527/acd89e}

\bibitem[{Z.-W. {Liu} {et~al.}(2015){Liu}, {Tauris}, {R{\"o}pke}, {Moriya}, {Kruckow}, {Stancliffe}, \& {Izzard}}]{Liu2015}
{Liu}, Z.-W., {Tauris}, T.~M., {R{\"o}pke}, F.~K., {et~al.} 2015, \bibinfo{title}{{The interaction of core-collapse supernova ejecta with a companion star},} \aap, 584, A11, \dodoi{10.1051/0004-6361/201526757}

\bibitem[{Z.-W. {Liu} {et~al.}(2013{\natexlab{b}}){Liu}, {Pakmor}, {Seitenzahl}, {Hillebrandt}, {Kromer}, {R{\"o}pke}, {Edelmann}, {Taubenberger}, {Maeda}, {Wang}, \& {Han}}]{Liu2013c}
{Liu}, Z.-W., {Pakmor}, R., {Seitenzahl}, I.~R., {et~al.} 2013{\natexlab{b}}, \bibinfo{title}{{The Impact of Type Ia Supernova Explosions on Helium Companions in the Chandrasekhar-mass Explosion Scenario},} \apj, 774, 37, \dodoi{10.1088/0004-637X/774/1/37}

\bibitem[{W. {Lu} {et~al.}(2025){Lu}, {Cary}, \& {Tsuna}}]{Lu2025}
{Lu}, W., {Cary}, S., \& {Tsuna}, D. 2025, \bibinfo{title}{{Accretion from a shock-inflated companion: double-peaked supernova lightcurve with periodic modulations},} arXiv e-prints, arXiv:2507.14284.
\newblock \doarXiv{2507.14284}

\bibitem[{L.~B. {Lucy}(1991){Lucy}}]{Lucy1991}
{Lucy}, L.~B. 1991, \bibinfo{title}{{Nonthermal Excitation of Helium in Type Ib Supernovae},} \apj, 383, 308, \dodoi{10.1086/170787}

\bibitem[{P. {Lundqvist} {et~al.}(2013){Lundqvist}, {Mattila}, {Sollerman}, {Kozma}, {Baron}, {Cox}, {Fransson}, {Leibundgut}, \& {Spyromilio}}]{Lundqvist2013}
{Lundqvist}, P., {Mattila}, S., {Sollerman}, J., {et~al.} 2013, \bibinfo{title}{{Hydrogen and helium in the spectra of Type Ia supernovae},} \mnras, 435, 329, \dodoi{10.1093/mnras/stt1303}

\bibitem[{P. {Lundqvist} {et~al.}(2015){Lundqvist}, {Nyholm}, {Taddia}, {Sollerman}, {Johansson}, {Kozma}, {Lundqvist}, {Fransson}, {Garnavich}, {Kromer}, {Shappee}, \& {Goobar}}]{Lundqvist2015}
{Lundqvist}, P., {Nyholm}, A., {Taddia}, F., {et~al.} 2015, \bibinfo{title}{{No trace of a single-degenerate companion in late spectra of supernovae 2011fe and 2014J},} \aap, 577, A39, \dodoi{10.1051/0004-6361/201525719}

\bibitem[{K. {Maguire} {et~al.}(2016){Maguire}, {Taubenberger}, {Sullivan}, \& {Mazzali}}]{Maguire2016}
{Maguire}, K., {Taubenberger}, S., {Sullivan}, M., \& {Mazzali}, P.~A. 2016, \bibinfo{title}{{Searching for swept-up hydrogen and helium in the late-time spectra of 11 nearby Type Ia supernovae},} \mnras, 457, 3254, \dodoi{10.1093/mnras/stv2991}

\bibitem[{E. {Marietta} {et~al.}(2000){Marietta}, {Burrows}, \& {Fryxell}}]{Marietta2000}
{Marietta}, E., {Burrows}, A., \& {Fryxell}, B. 2000, \bibinfo{title}{{Type IA Supernova Explosions in Binary Systems: The Impact on the Secondary Star and Its Consequences},} \apjs, 128, 615, \dodoi{10.1086/313392}

\bibitem[{S. {Mattila} {et~al.}(2005){Mattila}, {Lundqvist}, {Sollerman}, {Kozma}, {Baron}, {Fransson}, {Leibundgut}, \& {Nomoto}}]{Mattila2005}
{Mattila}, S., {Lundqvist}, P., {Sollerman}, J., {et~al.} 2005, \bibinfo{title}{{Early and late time VLT spectroscopy of SN 2001el - progenitor constraints for a type Ia supernova},} \aap, 443, 649, \dodoi{10.1051/0004-6361:20052731}

\bibitem[{C. {McCutcheon} {et~al.}(2022){McCutcheon}, {Zeng}, {Liu}, {Izzard}, {Pan}, {Chen}, \& {Han}}]{McCutcheon2022}
{McCutcheon}, C., {Zeng}, Y., {Liu}, Z.~W., {et~al.} 2022, \bibinfo{title}{{Type Ia supernova ejecta-donor interaction: explosion model comparison},} \mnras, 514, 4078, \dodoi{10.1093/mnras/stac1275}

\bibitem[{B. {Paczy{\'n}ski}(1971){Paczy{\'n}ski}}]{Paczynski1971}
{Paczy{\'n}ski}, B. 1971, \bibinfo{title}{{Evolutionary Processes in Close Binary Systems},} \araa, 9, 183, \dodoi{10.1146/annurev.aa.09.090171.001151}

\bibitem[{R. {Pakmor} {et~al.}(2008){Pakmor}, {R{\"o}pke}, {Weiss}, \& {Hillebrandt}}]{Pakmor2008}
{Pakmor}, R., {R{\"o}pke}, F.~K., {Weiss}, A., \& {Hillebrandt}, W. 2008, \bibinfo{title}{{The impact of type Ia supernovae on main sequence binary companions},} \aap, 489, 943, \dodoi{10.1051/0004-6361:200810456}

\bibitem[{R. {Pakmor} {et~al.}(2022){Pakmor}, {Callan}, {Collins}, {de Mink}, {Holas}, {Kerzendorf}, {Kromer}, {Neunteufel}, {O'Brien}, {R{\"o}pke}, {Ruiter}, {Seitenzahl}, {Shingles}, {Sim}, \& {Taubenberger}}]{Pakmor2022}
{Pakmor}, R., {Callan}, F.~P., {Collins}, C.~E., {et~al.} 2022, \bibinfo{title}{{On the fate of the secondary white dwarf in double-degenerate double-detonation Type Ia supernovae},} \mnras, 517, 5260, \dodoi{10.1093/mnras/stac3107}

\bibitem[{K.-C. {Pan} {et~al.}(2010){Pan}, {Ricker}, \& {Taam}}]{Pan2010}
{Pan}, K.-C., {Ricker}, P.~M., \& {Taam}, R.~E. 2010, \bibinfo{title}{{Impact of Type Ia Supernova Ejecta on a Helium-star Binary Companion},} \apj, 715, 78, \dodoi{10.1088/0004-637X/715/1/78}

\bibitem[{K.-C. {Pan} {et~al.}(2012){Pan}, {Ricker}, \& {Taam}}]{Pan2012a}
{Pan}, K.-C., {Ricker}, P.~M., \& {Taam}, R.~E. 2012, \bibinfo{title}{{Impact of Type Ia Supernova Ejecta on Binary Companions in the Single-degenerate Scenario},} \apj, 750, 151, \dodoi{10.1088/0004-637X/750/2/151}

\bibitem[{K.-C. {Pan} {et~al.}(2025){Pan}, {Ruiz-Lapuente}, \& {Gonz{\'a}lez Hern{\'a}ndez}}]{Pan2025}
{Pan}, K.-C., {Ruiz-Lapuente}, P., \& {Gonz{\'a}lez Hern{\'a}ndez}, J.~I. 2025, \bibinfo{title}{{Supernova Remnants with M dwarf surviving companions},} arXiv e-prints, arXiv:2507.01331, \dodoi{10.48550/arXiv.2507.01331}

\bibitem[{O. {Papish} {et~al.}(2015){Papish}, {Soker}, {Garc{\'\i}a-Berro}, \& {Aznar-Sigu{\'a}n}}]{Papish2015}
{Papish}, O., {Soker}, N., {Garc{\'\i}a-Berro}, E., \& {Aznar-Sigu{\'a}n}, G. 2015, \bibinfo{title}{{The response of a helium white dwarf to an exploding Type Ia supernova},} \mnras, 449, 942, \dodoi{10.1093/mnras/stv337}

\bibitem[{B. {Paxton} {et~al.}(2011){Paxton}, {Bildsten}, {Dotter}, {Herwig}, {Lesaffre}, \& {Timmes}}]{MESAI}
{Paxton}, B., {Bildsten}, L., {Dotter}, A., {et~al.} 2011, \bibinfo{title}{{Modules for Experiments in Stellar Astrophysics (MESA)},} \apjs, 192, 3, \dodoi{10.1088/0067-0049/192/1/3}

\bibitem[{B. {Paxton} {et~al.}(2013){Paxton}, {Cantiello}, {Arras}, {Bildsten}, {Brown}, {Dotter}, {Mankovich}, {Montgomery}, {Stello}, {Timmes}, \& {Townsend}}]{MESAII}
{Paxton}, B., {Cantiello}, M., {Arras}, P., {et~al.} 2013, \bibinfo{title}{{Modules for Experiments in Stellar Astrophysics (MESA): Planets, Oscillations, Rotation, and Massive Stars},} \apjs, 208, 4, \dodoi{10.1088/0067-0049/208/1/4}

\bibitem[{B. {Paxton} {et~al.}(2015){Paxton}, {Marchant}, {Schwab}, {Bauer}, {Bildsten}, {Cantiello}, {Dessart}, {Farmer}, {Hu}, {Langer}, {Townsend}, {Townsley}, \& {Timmes}}]{MESAIII}
{Paxton}, B., {Marchant}, P., {Schwab}, J., {et~al.} 2015, \bibinfo{title}{{Modules for Experiments in Stellar Astrophysics (MESA): Binaries, Pulsations, and Explosions},} \apjs, 220, 15, \dodoi{10.1088/0067-0049/220/1/15}

\bibitem[{B. {Paxton} {et~al.}(2018){Paxton}, {Schwab}, {Bauer}, {Bildsten}, {Blinnikov}, {Duffell}, {Farmer}, {Goldberg}, {Marchant}, {Sorokina}, {Thoul}, {Townsend}, \& {Timmes}}]{MESAIV}
{Paxton}, B., {Schwab}, J., {Bauer}, E.~B., {et~al.} 2018, \bibinfo{title}{{Modules for Experiments in Stellar Astrophysics (MESA): Convective Boundaries, Element Diffusion, and Massive Star Explosions},} \apjs, 234, 34, \dodoi{10.3847/1538-4365/aaa5a8}

\bibitem[{B. {Paxton} {et~al.}(2019){Paxton}, {Smolec}, {Schwab}, {Gautschy}, {Bildsten}, {Cantiello}, {Dotter}, {Farmer}, {Goldberg}, {Jermyn}, {Kanbur}, {Marchant}, {Thoul}, {Townsend}, {Wolf}, {Zhang}, \& {Timmes}}]{MESAV}
{Paxton}, B., {Smolec}, R., {Schwab}, J., {et~al.} 2019, \bibinfo{title}{{Modules for Experiments in Stellar Astrophysics (MESA): Pulsating Variable Stars, Rotation, Convective Boundaries, and Energy Conservation},} \apjs, 243, 10, \dodoi{10.3847/1538-4365/ab2241}

\bibitem[{F. P{\'e}rez \& B.~E. Granger(2007)P{\'e}rez \& Granger}]{perez_2007_aa}
P{\'e}rez, F., \& Granger, B.~E. 2007, \bibinfo{title}{{IPython}: a system for interactive scientific computing,} Computing in Science \& Engineering, 9, 21

\bibitem[{S. {Perlmutter} {et~al.}(1999){Perlmutter}, {Aldering}, {Goldhaber}, {Knop}, {Nugent}, {Castro}, {Deustua}, {Fabbro}, {Goobar}, {Groom}, {Hook}, {Kim}, {Kim}, {Lee}, {Nunes}, {Pain}, {Pennypacker}, {Quimby}, {Lidman}, {Ellis}, {Irwin}, {McMahon}, {Ruiz-Lapuente}, {Walton}, {Schaefer}, {Boyle}, {Filippenko}, {Matheson}, {Fruchter}, {Panagia}, {Newberg}, {Couch}, \& {Project}}]{Perlmutter1999}
{Perlmutter}, S., {Aldering}, G., {Goldhaber}, G., {et~al.} 1999, \bibinfo{title}{{Measurements of {\ensuremath{\Omega}} and {\ensuremath{\Lambda}} from 42 High-Redshift Supernovae},} \apj, 517, 565, \dodoi{10.1086/307221}

\bibitem[{A.~Y. {Potekhin} \& G. {Chabrier}(2010){Potekhin} \& {Chabrier}}]{Potekhin2010}
{Potekhin}, A.~Y., \& {Chabrier}, G. 2010, \bibinfo{title}{{Thermodynamic Functions of Dense Plasmas: Analytic Approximations for Astrophysical Applications},} Contributions to Plasma Physics, 50, 82, \dodoi{10.1002/ctpp.201010017}

\bibitem[{L.~J. {Prust} \& L. {Bildsten}(2024){Prust} \& {Bildsten}}]{Prust2024}
{Prust}, L.~J., \& {Bildsten}, L. 2024, \bibinfo{title}{{Flow morphology of a supersonic gravitating sphere},} \mnras, 527, 2869, \dodoi{10.1093/mnras/stad3405}

\bibitem[{L.~J. {Prust} {et~al.}(2025){Prust}, {Kumar}, \& {Bildsten}}]{Prust2025}
{Prust}, L.~J., {Kumar}, G., \& {Bildsten}, L. 2025, \bibinfo{title}{{Ejecta Wakes from Companion Interaction in Type Ia Supernova Remnants},} \apj, 982, 60, \dodoi{10.3847/1538-4357/adb7db}

\bibitem[{G. {Ramsay} {et~al.}(2018){Ramsay}, {Green}, {Marsh}, {Kupfer}, {Breedt}, {Korol}, {Groot}, {Knigge}, {Nelemans}, {Steeghs}, {Woudt}, \& {Aungwerojwit}}]{Ramsay2018}
{Ramsay}, G., {Green}, M.~J., {Marsh}, T.~R., {et~al.} 2018, \bibinfo{title}{{Physical properties of AM CVn stars: New insights from Gaia DR2},} \aap, 620, A141, \dodoi{10.1051/0004-6361/201834261}

\bibitem[{S.-J. {Rau} \& K.-C. {Pan}(2022){Rau} \& {Pan}}]{Rau2022}
{Rau}, S.-J., \& {Pan}, K.-C. 2022, \bibinfo{title}{{Evolution of Main-sequence-like Surviving Companions in Type Ia Supernova Remnants},} \apj, 933, 38, \dodoi{10.3847/1538-4357/ac7153}

\bibitem[{A.~G. {Riess} {et~al.}(1998){Riess}, {Filippenko}, {Challis}, {Clocchiatti}, {Diercks}, {Garnavich}, {Gilliland}, {Hogan}, {Jha}, {Kirshner}, {Leibundgut}, {Phillips}, {Reiss}, {Schmidt}, {Schommer}, {Smith}, {Spyromilio}, {Stubbs}, {Suntzeff}, \& {Tonry}}]{Riess1998}
{Riess}, A.~G., {Filippenko}, A.~V., {Challis}, P., {et~al.} 1998, \bibinfo{title}{{Observational Evidence from Supernovae for an Accelerating Universe and a Cosmological Constant},} \aj, 116, 1009, \dodoi{10.1086/300499}

\bibitem[{A. {Rimoldi} {et~al.}(2016){Rimoldi}, {Portegies Zwart}, \& {Rossi}}]{Rimoldi2016}
{Rimoldi}, A., {Portegies Zwart}, S., \& {Rossi}, E.~M. 2016, \bibinfo{title}{{Simulations of stripped core-collapse supernovae in close binaries},} Computational Astrophysics and Cosmology, 3, 2, \dodoi{10.1186/s40668-016-0015-4}

\bibitem[{F.~J. {Rogers} \& A. {Nayfonov}(2002){Rogers} \& {Nayfonov}}]{Rogers2002}
{Rogers}, F.~J., \& {Nayfonov}, A. 2002, \bibinfo{title}{{Updated and Expanded OPAL Equation-of-State Tables: Implications for Helioseismology},} \apj, 576, 1064, \dodoi{10.1086/341894}

\bibitem[{A.~J. {Ruiter} \& I.~R. {Seitenzahl}(2025){Ruiter} \& {Seitenzahl}}]{Ruiter2025}
{Ruiter}, A.~J., \& {Seitenzahl}, I.~R. 2025, \bibinfo{title}{{Type Ia supernova progenitors: a contemporary view of a long-standing puzzle},} \aapr, 33, 1, \dodoi{10.1007/s00159-024-00158-9}

\bibitem[{D.~J. {Sand} {et~al.}(2018){Sand}, {Graham}, {Boty{\'a}nszki}, {Hiramatsu}, {McCully}, {Valenti}, {Hosseinzadeh}, {Howell}, {Burke}, {Cartier}, {Diamond}, {Hsiao}, {Jha}, {Kasen}, {Kumar}, {Marion}, {Suntzeff}, {Tartaglia}, {Wheeler}, \& {Wyatt}}]{Sand2018}
{Sand}, D.~J., {Graham}, M.~L., {Boty{\'a}nszki}, J., {et~al.} 2018, \bibinfo{title}{{Nebular Spectroscopy of the {\textquotedblleft}Blue Bump{\textquotedblright} Type Ia Supernova 2017cbv},} \apj, 863, 24, \dodoi{10.3847/1538-4357/aacde8}

\bibitem[{D.~J. {Sand} {et~al.}(2021){Sand}, {Sarbadhicary}, {Pellegrino}, {Misra}, {Dastidar}, {Brown}, {Itagaki}, {Valenti}, {Swift}, {Andrews}, {Bostroem}, {Burke}, {Chomiuk}, {Dong}, {Galbany}, {Graham}, {Hiramatsu}, {Howell}, {Hsiao}, {Janzen}, {Jencson}, {Lundquist}, {McCully}, {Reichart}, {Smith}, {Wang}, \& {Wyatt}}]{Sand2021}
{Sand}, D.~J., {Sarbadhicary}, S.~K., {Pellegrino}, C., {et~al.} 2021, \bibinfo{title}{{Circumstellar Medium Constraints on the Environment of Two Nearby Type Ia Supernovae: SN 2017cbv and SN 2020nlb},} \apj, 922, 21, \dodoi{10.3847/1538-4357/ac20da}

\bibitem[{D. {Saumon} {et~al.}(1995){Saumon}, {Chabrier}, \& {van Horn}}]{Saumon1995}
{Saumon}, D., {Chabrier}, G., \& {van Horn}, H.~M. 1995, \bibinfo{title}{{An Equation of State for Low-Mass Stars and Giant Planets},} \apjs, 99, 713, \dodoi{10.1086/192204}

\bibitem[{I.~R. {Seitenzahl} {et~al.}(2013){Seitenzahl}, {Ciaraldi-Schoolmann}, {R{\"o}pke}, {Fink}, {Hillebrandt}, {Kromer}, {Pakmor}, {Ruiter}, {Sim}, \& {Taubenberger}}]{Seitenzahl2013}
{Seitenzahl}, I.~R., {Ciaraldi-Schoolmann}, F., {R{\"o}pke}, F.~K., {et~al.} 2013, \bibinfo{title}{{Three-dimensional delayed-detonation models with nucleosynthesis for Type Ia supernovae},} \mnras, 429, 1156, \dodoi{10.1093/mnras/sts402}

\bibitem[{B.~J. {Shappee} \& K.~Z. {Stanek}(2011){Shappee} \& {Stanek}}]{Shappee2011}
{Shappee}, B.~J., \& {Stanek}, K.~Z. 2011, \bibinfo{title}{{A New Cepheid Distance to the Giant Spiral M101 Based on Image Subtraction of Hubble Space Telescope/Advanced Camera for Surveys Observations},} \apj, 733, 124, \dodoi{10.1088/0004-637X/733/2/124}

\bibitem[{B.~J. {Shappee} {et~al.}(2017){Shappee}, {Stanek}, {Kochanek}, \& {Garnavich}}]{Shappee2017}
{Shappee}, B.~J., {Stanek}, K.~Z., {Kochanek}, C.~S., \& {Garnavich}, P.~M. 2017, \bibinfo{title}{{Whimper of a Bang: Documenting the Final Days of the Nearby Type Ia Supernova 2011fe},} \apj, 841, 48, \dodoi{10.3847/1538-4357/aa6eab}

\bibitem[{K.~J. {Shen}(2025){Shen}}]{Shen2025}
{Shen}, K.~J. 2025, \bibinfo{title}{{The Evolution of Hypervelocity Supernova Survivors and the Outcomes of Interacting Double White Dwarf Binaries},} \apj, 982, 6, \dodoi{10.3847/1538-4357/adb42e}

\bibitem[{K.~J. {Shen} {et~al.}(2024){Shen}, {Boos}, \& {Townsley}}]{Shen2024}
{Shen}, K.~J., {Boos}, S.~J., \& {Townsley}, D.~M. 2024, \bibinfo{title}{{Almost All Carbon/Oxygen White Dwarfs Can Support Double Detonations},} arXiv e-prints, arXiv:2405.19417, \dodoi{10.48550/arXiv.2405.19417}

\bibitem[{K.~J. {Shen} {et~al.}(2018{\natexlab{a}}){Shen}, {Kasen}, {Miles}, \& {Townsley}}]{Shen2018_subMch}
{Shen}, K.~J., {Kasen}, D., {Miles}, B.~J., \& {Townsley}, D.~M. 2018{\natexlab{a}}, \bibinfo{title}{{Sub-Chandrasekhar-mass White Dwarf Detonations Revisited},} \apj, 854, 52, \dodoi{10.3847/1538-4357/aaa8de}

\bibitem[{K.~J. {Shen} {et~al.}(2018{\natexlab{b}}){Shen}, {Boubert}, {G{\"a}nsicke}, {Jha}, {Andrews}, {Chomiuk}, {Foley}, {Fraser}, {Gromadzki}, {Guillochon}, {Kotze}, {Maguire}, {Siebert}, {Smith}, {Strader}, {Badenes}, {Kerzendorf}, {Koester}, {Kromer}, {Miles}, {Pakmor}, {Schwab}, {Toloza}, {Toonen}, {Townsley}, \& {Williams}}]{Shen2018_D6}
{Shen}, K.~J., {Boubert}, D., {G{\"a}nsicke}, B.~T., {et~al.} 2018{\natexlab{b}}, \bibinfo{title}{{Three Hypervelocity White Dwarfs in Gaia DR2: Evidence for Dynamically Driven Double-degenerate Double-detonation Type Ia Supernovae},} \apj, 865, 15, \dodoi{10.3847/1538-4357/aad55b}

\bibitem[{J.~V. {Shields} {et~al.}(2023){Shields}, {Arunachalam}, {Kerzendorf}, {Hughes}, {Biriouk}, {Monk}, \& {Buchner}}]{Shields2023}
{Shields}, J.~V., {Arunachalam}, P., {Kerzendorf}, W., {et~al.} 2023, \bibinfo{title}{{No Surviving SN Ia Companion in SNR 0509-67.5: Stellar Population Characterization and Comparison to Models},} \apjl, 950, L10, \dodoi{10.3847/2041-8213/acd6a0}

\bibitem[{J.~V. {Shields} {et~al.}(2022){Shields}, {Kerzendorf}, {Hosek}, {Shen}, {Rest}, {Do}, {Lu}, {Fullard}, {Strampelli}, \& {Zenteno}}]{Shields2022}
{Shields}, J.~V., {Kerzendorf}, W., {Hosek}, M.~W., {et~al.} 2022, \bibinfo{title}{{Searching for a Hypervelocity White Dwarf SN Ia Companion: A Proper-motion Survey of SN 1006},} \apjl, 933, L31, \dodoi{10.3847/2041-8213/ac7950}

\bibitem[{J.~M. {Stone} {et~al.}(2020){Stone}, {Tomida}, {White}, \& {Felker}}]{Stone2020}
{Stone}, J.~M., {Tomida}, K., {White}, C.~J., \& {Felker}, K.~G. 2020, \bibinfo{title}{{The Athena++ Adaptive Mesh Refinement Framework: Design and Magnetohydrodynamic Solvers},} \apjs, 249, 4, \dodoi{10.3847/1538-4365/ab929b}

\bibitem[{ {STScI Development Team}(2020){STScI Development Team}}]{stsynphot}
{STScI Development Team}. 2020, \bibinfo{title}{{stsynphot: synphot for HST and JWST},}, Astrophysics Source Code Library, record ascl:2010.003

\bibitem[{A. {Tanikawa} {et~al.}(2019){Tanikawa}, {Nomoto}, {Nakasato}, \& {Maeda}}]{Tanikawa2019}
{Tanikawa}, A., {Nomoto}, K., {Nakasato}, N., \& {Maeda}, K. 2019, \bibinfo{title}{{Double-detonation Models for Type Ia Supernovae: Trigger of Detonation in Companion White Dwarfs and Signatures of Companions{\textquoteright} Stripped-off Materials},} \apj, 885, 103, \dodoi{10.3847/1538-4357/ab46b6}

\bibitem[{F.~X. {Timmes} \& F.~D. {Swesty}(2000){Timmes} \& {Swesty}}]{Timmes2000}
{Timmes}, F.~X., \& {Swesty}, F.~D. 2000, \bibinfo{title}{{The Accuracy, Consistency, and Speed of an Electron-Positron Equation of State Based on Table Interpolation of the Helmholtz Free Energy},} \apjs, 126, 501, \dodoi{10.1086/313304}

\bibitem[{K. {Tomida} \& J.~M. {Stone}(2023){Tomida} \& {Stone}}]{Tomida2023}
{Tomida}, K., \& {Stone}, J.~M. 2023, \bibinfo{title}{{The Athena++ Adaptive Mesh Refinement Framework: Multigrid Solvers for Self-gravity},} \apjs, 266, 7, \dodoi{10.3847/1538-4365/acc2c0}

\bibitem[{M.~A. {Tucker} \& B.~J. {Shappee}(2024){Tucker} \& {Shappee}}]{Tucker2024}
{Tucker}, M.~A., \& {Shappee}, B.~J. 2024, \bibinfo{title}{{The HST Nondetection of SN Ia 2011fe 11.5 yr after Explosion Further Restricts Single-degenerate Progenitor Systems},} \apj, 962, 74, \dodoi{10.3847/1538-4357/ad1b4e}

\bibitem[{M.~A. {Tucker} {et~al.}(2022){Tucker}, {Shappee}, {Kochanek}, {Stanek}, {Ashall}, {Anand}, \& {Garnavich}}]{Tucker2022}
{Tucker}, M.~A., {Shappee}, B.~J., {Kochanek}, C.~S., {et~al.} 2022, \bibinfo{title}{{The whisper of a whimper of a bang: 2400 d of the Type Ia SN 2011fe reveals the decay of $^{55}$Fe},} \mnras, 517, 4119, \dodoi{10.1093/mnras/stac2873}

\bibitem[{M.~A. {Tucker} {et~al.}(2019){Tucker}, {Shappee}, \& {Wisniewski}}]{Tucker2019}
{Tucker}, M.~A., {Shappee}, B.~J., \& {Wisniewski}, J.~P. 2019, \bibinfo{title}{{No Stripped Companion Material in the Nebular Spectrum of the {\textquotedblleft}Two-Component{\textquotedblright} Type Ia Supernova ASASSN-18bt},} \apjl, 872, L22, \dodoi{10.3847/2041-8213/ab0286}

\bibitem[{M.~A. {Tucker} {et~al.}(2020){Tucker}, {Shappee}, {Vallely}, {Stanek}, {Prieto}, {Botyanszki}, {Kochanek}, {Anderson}, {Brown}, {Galbany}, {Holoien}, {Hsiao}, {Kumar}, {Kuncarayakti}, {Morrell}, {Phillips}, {Stritzinger}, \& {Thompson}}]{Tucker2020}
{Tucker}, M.~A., {Shappee}, B.~J., {Vallely}, P.~J., {et~al.} 2020, \bibinfo{title}{{Nebular spectra of 111 Type Ia supernovae disfavour single-degenerate progenitors},} \mnras, 493, 1044, \dodoi{10.1093/mnras/stz3390}

\bibitem[{B. {van Leer}(1979){van Leer}}]{vanLeer1979}
{van Leer}, B. 1979, \bibinfo{title}{{Towards the Ultimate Conservative Difference Scheme. V. A Second-Order Sequel to Godunov's Method},} Journal of Computational Physics, 32, 101, \dodoi{10.1016/0021-9991(79)90145-1}

\bibitem[{J. {van Roestel} {et~al.}(2022){van Roestel}, {Kupfer}, {Green}, {Wong}, {Bildsten}, {Burdge}, {Prince}, {Marsh}, {Szkody}, {Fremling}, {Graham}, {Dhillon}, {Littlefair}, {Bellm}, {Coughlin}, {Duev}, {Goldstein}, {Laher}, {Rusholme}, {Riddle}, {Dekany}, \& {Kulkarni}}]{vanRoestel2022}
{van Roestel}, J., {Kupfer}, T., {Green}, M.~J., {et~al.} 2022, \bibinfo{title}{{Discovery and characterization of five new eclipsing AM CVn systems},} \mnras, 512, 5440, \dodoi{10.1093/mnras/stab2421}

\bibitem[{P. {Virtanen} {et~al.}(2020){Virtanen}, {Gommers}, {Oliphant}, {Haberland}, {Reddy}, {Cournapeau}, {Burovski}, {Peterson}, {Weckesser}, {Bright}, {van der Walt}, {Brett}, {Wilson}, {Millman}, {Mayorov}, {Nelson}, {Jones}, {Kern}, {Larson}, {Carey}, {Polat}, {Feng}, {Moore}, {VanderPlas}, {Laxalde}, {Perktold}, {Cimrman}, {Henriksen}, {Quintero}, {Harris}, {Archibald}, {Ribeiro}, {Pedregosa}, {van Mulbregt}, \& {SciPy 1. 0 Contributors}}]{scipy2020}
{Virtanen}, P., {Gommers}, R., {Oliphant}, T.~E., {et~al.} 2020, \bibinfo{title}{{SciPy 1.0: fundamental algorithms for scientific computing in Python},} Nature Methods, 17, 261, \dodoi{10.1038/s41592-019-0686-2}

\bibitem[{K. {Werner} {et~al.}(2024){Werner}, {Reindl}, {Rauch}, {El-Badry}, \& {B{\'e}dard}}]{Werner2024}
{Werner}, K., {Reindl}, N., {Rauch}, T., {El-Badry}, K., \& {B{\'e}dard}, A. 2024, \bibinfo{title}{{The photospheres of the hottest fastest stars in the Galaxy},} \aap, 682, A42, \dodoi{10.1051/0004-6361/202348286}

\bibitem[{J.~C. {Wheeler} {et~al.}(1975){Wheeler}, {Lecar}, \& {McKee}}]{Wheeler1975}
{Wheeler}, J.~C., {Lecar}, M., \& {McKee}, C.~F. 1975, \bibinfo{title}{{Supernovae in binary systems.},} \apj, 200, 145, \dodoi{10.1086/153771}

\bibitem[{B. Wolf \& J. Schwab(2017)Wolf \& Schwab}]{bill_wolf_2017_826958}
Wolf, B., \& Schwab, J. 2017, \bibinfo{title}{wmwolf/py\_mesa\_reader: Interact with MESA Output,}, 0.3.0 Zenodo, \dodoi{10.5281/zenodo.826958}

\bibitem[{T.~L.~S. {Wong} \& L. {Bildsten}(2023){Wong} \& {Bildsten}}]{Wong2023}
{Wong}, T. L.~S., \& {Bildsten}, L. 2023, \bibinfo{title}{{Dynamical He Flashes in Double White Dwarf Binaries},} \apj, 951, 28, \dodoi{10.3847/1538-4357/acce9d}

\bibitem[{T.~L.~S. {Wong} {et~al.}(2024){Wong}, {White}, \& {Bildsten}}]{Wong2024}
{Wong}, T. L.~S., {White}, C.~J., \& {Bildsten}, L. 2024, \bibinfo{title}{{Shocking and Mass Loss of Compact Donor Stars in Type Ia Supernovae},} \apj, 973, 65, \dodoi{10.3847/1538-4357/ad6a11}

\bibitem[{N. {Yamaguchi} {et~al.}(2025){Yamaguchi}, {El-Badry}, {Wong}, \& {Shen}}]{Yamaguchi2025}
{Yamaguchi}, N., {El-Badry}, K., {Wong}, T. L.~S., \& {Shen}, K.~J. 2025, \bibinfo{title}{{Carbon burning cannot explain puffy hypervelocity white dwarfs},} arXiv e-prints, arXiv:2507.15952.
\newblock \doarXiv{2507.15952}

\bibitem[{P.~Z. {Yao} {et~al.}(2025){Yao}, {Quataert}, {Jiang}, {Lu}, \& {White}}]{Yao2025}
{Yao}, P.~Z., {Quataert}, E., {Jiang}, Y.-F., {Lu}, W., \& {White}, C.~J. 2025, \bibinfo{title}{{Star‑Disk Collisions: Implications for Quasi-periodic Eruptions and Other Transients near Supermassive Black Holes},} \apj, 978, 91, \dodoi{10.3847/1538-4357/ad8911}

\bibitem[{Y. {Zeng} {et~al.}(2020){Zeng}, {Liu}, \& {Han}}]{Zeng2020}
{Zeng}, Y., {Liu}, Z.-W., \& {Han}, Z. 2020, \bibinfo{title}{{The Interaction of Type Iax Supernova Ejecta with a Helium Companion Star},} \apj, 898, 12, \dodoi{10.3847/1538-4357/ab9943}

\bibitem[{M. {Zhang} {et~al.}(2019){Zhang}, {Fuller}, {Schwab}, \& {Foley}}]{Zhang2019}
{Zhang}, M., {Fuller}, J., {Schwab}, J., \& {Foley}, R.~J. 2019, \bibinfo{title}{{The Long-term Evolution and Appearance of Type Iax Postgenitor Stars},} \apj, 872, 29, \dodoi{10.3847/1538-4357/aafb34}

\end{thebibliography}
\bibliographystyle{aasjournalv7}

\end{document}